\documentclass[namedreferences,hyperref,optionalrh]{spr-sola}
\usepackage{graphicx}        
\usepackage{color}           
\usepackage{amsmath}

\chardef\us=`\_

\begin{document}

\begin{frontmatter}
\title{The \textit{Spectro-Polarimeter} (SP) of the \textit{Andrei B. Severny Solar Tower Telescope} (STT) at the \textit{Crimean Astrophysical Observatory}: Optical Design and Implementation}

\author[corref, email={alex.s.kutsenko@gmail.com}]{\inits{A.S.}\fnm{Alexander}~\snm{Kutsenko}\orcid{0000-0002-1196-5049}}
\author[email={valery@terebizh.ru}]{\inits{V.Yu.}\fnm{Valery}~\snm{Terebizh}\orcid{0000-0003-4129-3770}}
\author[email={dolgopolov.andrei@craocrimea.ru}]{\inits{A.V.}\fnm{Andrei}~\snm{Dolgopolov}\orcid{0000-0002-6494-3361}}
\author[email={vabramenko@gmail.com}]{\inits{V.I.}\fnm{Valentina}~\snm{Abramenko}\orcid{0000-0001-6466-4226}}
\author[email={plotnikov.andrey.alex@yandex.ru }]{\inits{A.A.}\fnm{Andrei}~\snm{Plotnikov}\orcid{0000-0001-8201-8391}}
\author{\inits{D.G.}\fnm{Dmitriy}~\snm{Semyonov}}
\author{\inits{V.N.}\fnm{Vladimir}~\snm{Skiruta}}
\author{\inits{V.I.}\fnm{Vyacheslav}~\snm{Lopukhin}}

\address{Crimean Astrophysical Observatory, Nauchny, 298409, Republic of Crimea}

\runningauthor{Kutsenko et al.}
\runningtitle{The \textit{Spectro-Polarimeter} of the \textit{Solar Tower Telescope}}

\begin{abstract}

The \textit{Spectro-Polarimeter} (SP) is a new instrument installed at the upgraded \textit{Andrei B. Severny Solar Tower Telescope} (STT) at the \textit{Crimean Astrophysical Observatory}. The instrument is a traditional echelle slit dual-beam spectropolarimeter with temporal modulation of the polarization. STT-SP provides simultaneous spectropolarimetric observations of the Sun within three 15~\AA\ wide spectral ranges around photospheric Fe~I 5250~\AA, Fe~I 5324~\AA, and chromospheric Mg~I b2 5172~\AA\ spectral lines. The spectral resolution of the instrument reaches 70,000 with the seeing-constrained slit width of 1 arcsec. The field-of-view of STT-SP is 200 arcsec allowing one to map a moderate size active region within a single raster scan. The instrument will provide new opportunities in the analysis of magnetic fields and thermodynamics of the lower atmosphere of the Sun. In this paper we describe the optical design of STT-SP and present the preliminary results acquired during the commissioning of the instrument.

\end{abstract}
\end{frontmatter}

\section{Introduction}
     \label{S-Introduction} 

Recent progress in the instrumentation for the observations of the Sun is really impressing. Modern facilities provide high spatial and temporal resolution solar data within the entire electromagnetic spectrum from radio to hard X-ray wavelengths. The new generation of optical solar telescopes such as the  \textit{Daniel K. Inouye Solar Telescope} \citep{Rimmele2020}, the \textit{Goode Solar Telescope} \citep{Cao2010}, and others allow one to explore solar features with the size of several tens of kilometers with unprecedented details. 

The essential requirement for nowadays solar optical instruments is the ability to perform multi-line high-precision spectropolarimetry. Thus, DKIST is equipped with three high-resolution spectropolarimeters covering together the spectral range from near-UV to mid-IR \citep{deWijn2022, Jaeggli2022, Fehlmann2023}. Two of the instruments, namely the \textit{Visible Spectro-Polarimeter} \citep{deWijn2022} and the \textit{Cryogenic Near-Infrared Spectro-Polarimeter} \citep{Fehlmann2023}, are able to observe the Sun within three different wavelength regions simultaneously providing a great flexibility for addressing numerous scientific issues. Several instruments of the solar telescope cluster at the Canary Islands are also equipped with high-class spectropolarimeters \citep[\textit{e.g.}][]{Scharmer2008, Collados2012}. The future \textit{European Solar Telescope} \citep{Collados2013} is designed to ensure the most accurate polarimetric measurements using the next-generation integral field spectropolarimeters \citep{QuinteroNoda2022}. The upgrade of existing instruments is also taking place in order to provide solar spectropolarimetry \citep[\textit{e.g.}][]{Pruthvi2023}.

The \textit{Andrei B. Severnyi Solar Tower Telescope} (STT) at the \textit{Crimean Astrophysical Observatory} (CrAO) is the largest optical solar telescope in Russia so far. The telescope was initially built by 1955 \citep{Severny1955} and had a 0.6 m primary mirror. By 1973, a major upgrade of the telescope has been finished \citep{Kotov1982}: the size of the optics and the focal length was increased significantly. Nowadays the telescope is an off-axis f/45 Nasmyth-Cassegrain with 0.9 m primary mirror. The size of the solar image in the focal plane is of about 480 mm with the image scale of 244 $\mu$m arcsec$^{-1}$. The telescope is fed by a pair of 1.2 m coelostat and 1.1 m auxiliary flat mirrors installed at the top of a 20 m-high tower. Due to off-axis construction the diffraction limit is not achievable: our analysis of the telescope model resulted in 0.3 arcsec theoretical spatial resolution. Our recent experiments on the speckle reconstruction of solar images acquired by STT \citep{Kutsenko2022} supported this deduction. Hence, despite more than 50 year age, the telescope optics and mechanical parts are still in an excellent condition.

The first-light instrument of STT was a dual-channel Babcock-type magnetograph \citep{Babcock1953}. The instrument was able to measure magnetic-field vector and Doppler velocity using two selected spectral lines. The spectral lines were recorded at two wavelength positions. The maps of the solar magnetic field were acquired by consecutive scanning of the solar surface through the magnetograph slit. Since 1990-s STT was used predominantly to probe the magnetic field and global oscillations of the Sun-as-a-star \citep{Haneychuk2003}. This kind of observations required the parallel rays from the Sun to be directed to the magnetograph slit by flat coelostat and folding mirrors. In other words, in fact, the telescope itself was not used to maintain that scientific program.

In 2018 our proposal on STT upgrade was supported by the Russian Science Foundation (Project 18-12-00131). Due to limited funds, we had to carefully analyze the concept of the telescope upgrade to make it as cost-effective as possible and to meet our scientific requirements. The installation of an adaptive optics (except for a zero-order one) was rejected. Instead, we intended to upgrade the control system of the telescope and to replace the magnetograph with a new spectropolarimeter. Our goal was to be able to carry out strictly simultaneous spectropolarimetric observations of several spectral line profiles formed within different height ranges. These data are supposed to be used to derive the 3D distribution of the magnetic field vector and Doppler velocities within lower layers of the solar atmosphere. With these observations, we are particular interested in measuring the tensor of current helicity and the full vector of electric current responsible for the storage of the energy released during solar flares. Since electric current relies on the rotor of magnetic field, usually only a $z$-related part of this quantity can be derived using magnetic field measurements at a single height. Although recent advances in Stokes profiles analysis enable to overcome this limitations \citep{PastorYabar2021}, measurements of the photospheric and chromospheric magnetic field vector might provide more direct way of electric current vector estimation. Multi-height magnetic field data also ensure more reliable coronal magnetic field extrapolation as compared to that derived from photopsheric magnetic field as a boundary conditions \citep[\textit{e.g.}][]{Jarolim2024}. Spatial data cubes on magnetic field and plasma velocity are essential for the analysis of MHD oscillations and energy propagation from the lower to upper atmosphere. More detailed description of the importance of chromospheric magnetic field measurement can be found, for instance, in a review by \citet{Lagg2017}.

In this paper we describe the newly created \textit{Spectro-Polarimeter} installed at STT. The instrument is a dual-beam slit echelle spectrograph with the polarization analyzer. SP provides full Stokes vector measurements within three spectral ranges in the vicinity of chromospheric Mg~I $b$ triplet and photospheric Fe~I 5250/5247~\AA\ lines. The details of the optical design and the first results are presented in the next sections.

\section{Design Concept and Requirements}
     \label{S-requirements} 

The first step in the development of a new instrument was to select the design concept of SP. Since tunable narrow-band filters were unavailable, we selected a classical slit spectrograph with the polarization analyzer. A tunable over a wide range spectrograph with expensive broadband camera and polarization optics was not an option as well. We were also limited to a single spectral arm due to cost efficiency. Hence, the final decision was a high-spectral-resolution spectrograph with the minimum of required elements operating within a certain wavelength range. The field-of-view of the instrument had to be comparable with the typical size of a moderate to large active region, \textit{i.e.} of about 150 arcsec. Taking into account the telescope focal length, pixel size of the detector, the desirable slit width of about 1 arcsec, and the spectral resolution of at least $R\approx 50,000$, our preliminary calculations limited the spectral range to about 50~\AA\ in the visible or near-IR part of the spectrum. 

Next we had to select the spectral range of the instrument. The number of chromospheric spectral lines within the visible and near-IR part of the spectrum is not so high \citep[see, \textit{e.g.}, section~5.3.1 in][]{Lagg2017}. The thermodynamical conditions in the chromosphere change dramatically with height resulting in non-LTE formation of spectral lines. Only the core of a spectral lines is formed in the chromosphere while the wings imprint the information of photospheric heights. Therefore, in principle, the observation of a single chromospheric line is sufficient to probe the magnetic field within both photosphere and chromosphere. However, the inversion of the Stokes profiles is more robust when an additional information on photospheric lines is included. Hence, our another requirement was the presence of a magnetically sensitive photospheric line in the close vicinity of the chromospheric one. The often used combination of chromospheric He~I 10830~\AA\ triplet and photospheric Si~I 10827~\AA\ line was rejected due to high cost of near-IR detectors. A well-suited for chromospheric observations Ca~II 8542~\AA\ line \citep[\textit{e.g.}][]{Cauzzi2008, DiazBaso2019} has no appropriate photopsheric companion within the estimated spectral range of our instrument. The use of excellent photospheric Fe~I 6301/6302~\AA\ spectral lines in conjunction with Ca~II 8542~\AA\ \citep[\textit{e.g. SOLIS/VSM}][]{Keller2003} would require a considerable complication of the optical layout.

After some discussion, we stopped our choice on Fraunhofer Mg~I $b$ triplet formed in the lower chromosphere. The properties of the triplet in regard of solar spectropolarimetry were analyzed recently by \citet{QuinteroNoda2018}. The Mg~I $b4$ line located at 5167~\AA\ is blended with magnetically sensitive photopsheric Fe~I line making the former inappropriate for inversions. Mg~I $b1$ and $b2$ located at 5184~\AA\ and 5172~\AA\ respectively are formed within almost the same height range and exhibit similar core intensity and line width, while the latter produces stronger response on magnetic field due to higher effective Land\'e factor $\overline{g}=1.75$. Hence, Mg~I $b2$ line widely used in solar observations is our target chromospheric line. According to \citet{QuinteroNoda2018}, the line core is formed predominantly at heights between 500 and 800 km allowing us to probe the magnetic field and plasma properties of the lower chromosphere. For comparison, the formation height of Ca~II 8542~\AA\ line lies mostly above 1000 km.

An obvious choice of the target photospheric spectral line in the vicinity of Mg~I b triplet is the Fe~I 5250.2~\AA\ spectral line. With the effective Land\'e factor $\overline{g}=3.00$, the line is one of the most sensitive to magnetic field line in this part of the visible spectrum. Beside, the Fe~I 5247/5250~\AA\ line pair provides additional diagnostic capabilities when used in magnetic line ratio approach \citep[see][and references therein]{Smitha2017}.

\section{The Upgrade of the Telescope Control System}
     \label{S-telescope} 

Before starting the telescope control system upgrade, we were inspired by the concept of the solar telescope control system presented in \citet{Varsik2006} and \citet{Yang2006}. We required the new control system to provide computer and manual control of the telescope pointing to the Sun and tracking according to the desired observational mode. The system must be distributed, \textit{i.e.} each task is primarily controlled by its own control computer or microprocessor. The communication between the parts is realized via the common core, which also provides GUI for the observer. Each functional part of the telescope must also be independent in the sense that the failure of this part does not affect the functionality of other units. Such a structure ensures easy future addition of new or upgrade of existing parts of the telescope control system.

The current structure of the telescope control system includes the pointing and guiding systems, the handling of the Spectro-Polarimeter and of the context imager. In this section we briefly describe the new pointing and guiding systems. A more detailed description will be published elsewhere.

\subsection{The Pointing System}
     \label{S-pointing} 
     
Since STT is a tower telescope, the pointing of the instrument to the desired point of the sky is carried out by rotating the coelostat (right ascension) and auxiliary (declination) mirrors \citep[\textit{e.g.}][]{Pasachoff1984}. In the previous control system built in 1973, the fine and coarse pointing in each coordinate was realized by several synchronous AC motors with different gear ratios. The pointing system was manually controlled by four consoles located within the telescope building. The computer control was not supported. Any modification of the pointing system required sophisticated hardware upgrades. It was practically impossible to reliably synchronize the telescope pointing and tracking with SP for scientific observations. This issue motivated us to implement a new pointing system.

First of all, we replaced all the AC motors by stepper motors in mirror mount drives. Since stepper motors provide flexibility in the rotation rate control without the torque loss, this decision led to decrease of the total number of drives from five to three. The drives are managed by an 8-axis stepper motor controller. The communication with the controller is realized via private Ethernet. The controller supports several simultaneous connections enabling to control the telescope by several hosts. Two microcomputer-based touchscreen consoles are installed at the top and at the bottom (near the focal plane of the telescope) of the tower to manually point the telescope to the desired part of the solar surface. The consoles provide user-friendly graphical user interface and communicate directly with the stepper motor controller.

The stepper motor controller is also able to control (i) the telescope dome rotation and opening, (ii) the movement of the coelostat mirror mount, and (iii) the focusing of the telescope by displacement of the secondary mirror. With the new pointing system, a control of the telescope by means of hand-held devices or through a web-based interface can be easily implemented. Any desired modifications require mostly software or firmware updates rather than hardware changes.

\subsection{The Guiding System}
     \label{S-guiding} 
     
STT images a part of the solar surface rather than the entire solar disk. Therefore, the image in the focal plane is unsuitable for guiding. For guiding and locating the region of interest we use an upgraded classical offset guider. The guider is a small refractor that forms a 180-mm in diameter white-light solar image on a vertical stage with moving circular carriage. For better accuracy, the guider is fed by a paraxial beam reflected by a plane 290-mm mirror installed at the center of the primary mirror of STT. Four photodetectors fixed within the moving carriage measure the light intensity at the solar limb. The photodetectors are located in opposite sides of the carriage roughly in the solar North-South and East-West directions. The solar light intensity is sampled by an analog-to-digital converter at a firmware defined frequency of hundreds of Hz. The samples are averaged by an on-board microprocessor and transferred to the main observer computer via private Ethernet. The computer software calculates the offset between the limb intensity measured by the pairs of the photodetectors and modify the slopes of the coelostat and auxiliary mirrors to minimize the offset. For this precision guiding, the coelostat and auxiliary mirror mounts can be additionally rotated within certain limits by linear actuators.

The carriage can be moved over the vertical stage in $X-Y$ directions. In general, these directions do not perfectly correspond to the celestial right ascension and declination directions. The observer computer software controls the travel of the carriage and, as a consequence, the displacement of the solar image in the focal plane of STT. During the observations, the solar image must move perpendicular to the spectrograph slit, hence the guider carriage has to move along an inclined trajectory with respect to its $X-Y$ coordinates. To find this trajectory parameters, a guider calibration routine is carried out daily prior to observations. During the routine, the carriage follows the image of the Sun. The observer computer software measures the consecutive positions of the solar image in stage $X-Y$ coordinates while the telescope moves along right ascension direction. These data are then approximated by a linear fit. The fit coefficients are used to calculate the positions of the carriage. The accuracy of the carriage positioning corresponds to 0.05 arcsec.

A software defined guider provides inaccessible earlier flexibility in the telescope control. For instance, different trajectories and velocities of the solar image motion over the spectrograph slit can be easily implemented. The guider can be used to estimate the total solar intensity (varying significantly during a day due to solar declination) for further photometric calibration of the spectral data. The automatic stop of the guiding is also realized when the intensity falls below certain threshold due to clouds. Finally, the guider can be easily integrated in the future tip-tilt system for better positioning and image stabilization.

\section{Instrument Design}
     \label{S-design} 

\begin{figure}    
\centerline{\includegraphics[width=1.\textwidth,clip=]{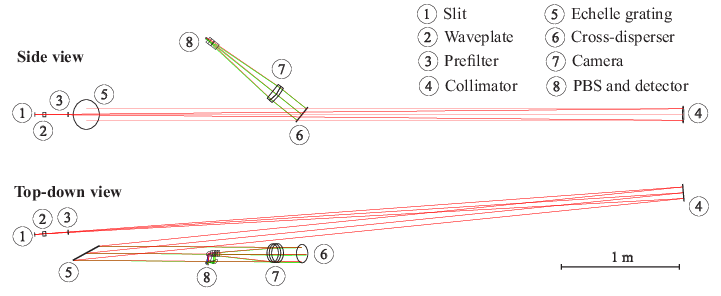}}
\small
        \caption{The side view (top panel) and top-down view (bottom panel) of the optical layout of SP. The components are labelled with numbers corresponding to the order of light beam propagation.}
\label{F-optical_layout}
\end{figure}

Beside scientific requirements, the main challenge in the designing of SP was to keep it as cost effective as possible. Therefore we were forced to avoid fabrication of custom optical elements and to utilize stock-produced items. The second challenge was to fit the slow long-focus telescope to modern detectors while keeping an appropriate field-of-view. The starting point was to measure the geometry of the telescope and of the solar image in the focal plane. These data were used to create an optical model of STT. The further design of SP was carried out in combination with the telescope optical layout. The ray-trace diagram of SP is provided in Figure~\ref{F-optical_layout}. 

\subsection{Slit and Context Imager}
     \label{S-slit} 
     
SP is installed in a separate room divided from the telescope tower by a steel 1-m in diameter disk. The mirror slit (item 1 in Figure~\ref{F-optical_layout}) is mounted on this disk in the focal plane of the telescope. The centers of auxiliary, primary, secondary mirrors and of the slit are located precisely in the plane of the local meridian. As a consequence, the rotation of the coelostat mirror (right ascension) results in the movement of the solar image along the horizontal direction in the focal plane. The axis of the Sun in the focal plane solar image is rotated by the $p$-angle from the vertical direction, \textit{i.e.} the solar North pole is roughly at the top of the image. Solar active regions are usually oriented in the East-West direction. Hence, the slit is vertically oriented allowing us to capture the entire active region of interest with a single field scan in the direction perpendicular to the slit height. The scanning is achieved by rotating the coelostat at a desired rate. 

The slit is made of two polished moving stainless-steel plates. The width of the slit is finely adjusted by a micrometer screw between 0 and 2 mm. The typical used slit width is 250~$\mu$m corresponding to approximately 1 arcsec at 525 nm. The slit height is 70 mm. However, the field-of-view is limited by the 48-mm clear aperture polarization optics installed immediately after the slit (item 2 in Figure~\ref{F-optical_layout}).

The slit is slightly inclined with respect to the telescope focal plane to reflect an incoming beam to the context imager (not shown in Figure~\ref{F-optical_layout}). The context imager consists of a folding mirror, directing the beam along the vertical tower of the telescope, and two lenses. The lenses re-focus the solar light at a 2048$\times$2048 pixel detector resulting in 150$\times$150 arcsec$^2$ spatial field-of-view. On its way, the light transmits through a 10~\AA\ FWHM interference filter centered at 7057~\AA\ TiO molecular band. The detector takes the images at a 75 frames per second rate.

\subsection{Spectrograph}
     \label{S-spectrograph} 

The spectral resolution of a grating spectrograph can be evaluated as (V.~Terebizh, 2024, private communication)
\begin{equation}  \label{Eq-resolution}
 R = \frac{2\ \mathrm{sin}\delta \ \mathrm{cos}(\alpha-\delta)}{\mathrm{cos}\alpha} \frac{\Delta}{\omega D},
\end{equation}
where $\alpha$ is the angle of incidence on the diffraction grating, $\delta$ is the blaze angle of the diffraction grating, $\Delta$ is the diameter of the collimated beam, $D$ is the diameter of the primary mirror, and $\omega$ is the angular size of the slit in radians. The angular size of the slit is limited by seeing to about 1 arcsec. The main parameters we can vary to achieve the highest spectral resolution is the diameter of the collimated beam and the blaze angle of the grating. The blaze angle, in turn, is defined by the optimal angular dispersion. In our case, the largest diameter of the collimated beam, $\Delta$, was driven by the size of the available diffraction grating and by the height of the slit. The chosen grating was an 120$\times$250 mm$^2$ echelle, which constrain the size of the rest optical elements of the spectrograph.

The spectrograph consists of a collimator mirror (item 4 in Figure~\ref{F-optical_layout}), echelle diffraction grating (item 5 in Figure~\ref{F-optical_layout}), cross-disperser, and a camera doublet (items 6 and 7 in Figure~\ref{F-optical_layout}, respectively). All the elements except for the camera, detector, and polarization beamsplitter, lie in a horizontal plane as can be seen from Figure~\ref{F-optical_layout}. We further discuss each of the elements in more detail.

\begin{figure}    
\centerline{\includegraphics[width=0.4\textwidth,clip=]{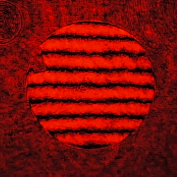}}
\small
        \caption{
        The interferogram of the collimator mirror made to test the quality of the surface. The monochromatic light source is a He-Ne laser at 6328~\AA. The distortion of the wavefront does not exceed $\lambda/10$ within the most part of the mirror area.
        }
\label{F-interferogram}
\end{figure}

The collimator (item 4 in Figure~\ref{F-optical_layout}) is a 140 mm in diameter spherical mirror with the focal length of 5415 mm, which was driven by the size and tilt angle of the diffraction grating. Due to large focal ratio $\approx f/38$ and small angle between the incoming and outgoing beams (2 degree) the spherical aberrations of the collimator are negligible. The collimator was fabricated of glass-ceramic at the Crimean Astrophysical Observatory. The mirror is coated with aluminium. An excellent quality of the surface can be seen in Figure~\ref{F-interferogram}. The figure demonstrates the interferogram of the mirror made with He-Ne laser at 6328~\AA. The interference pattern reveals the wavefront deviation to be mostly not more than $\lambda/10$. The collimator mount provides all kind of fine adjustment of the mirror.

The collimator images the collimated beam to the diffraction grating (item 5 in Figure~\ref{F-optical_layout}). The diffraction grating is an R2 aluminium coated echelle with the blaze angle of 64 degree. The groove density is 50 mm$^{-1}$. Both incoming and outgoing beams lie in the horizontal plane with the separation angle 8.5 degree. The diffraction grating was mounted on the steel disk that separates the telescope tower from the spectrograph. The mount of the grating provides fine manual adjustment of the grating normal with respect to the incoming beam in the horizontal plane. Hence, with different camera, we are able to observe the Sun within any part of the visible spectrum.

With the selected spectrograph geometry, the target wavelength ranges around Mg~I 5172~\AA\ and Fe~I 5250~\AA\ spectral lines are located near the centers of the overlapping 66$^{th}$ and 65$^{th}$ spectral orders of the grating, respectively. To spatially separate the orders, we use a flat diffraction grating with 600 grooves~mm$^{-1}$ as a cross-disperser (item 6 in Figure~\ref{F-optical_layout}). The grating is used in the fifth order which provides the optimum value for the required angular dispersion across the main dispersion of the echelle grating. The cross-disperser is also optimized to provide the maximum outgoing light intensity in the fifth order. The grating was retrieved from the old magnetograph and the size of the grating is excessive: with the size of 300$\times$200~mm$^{2}$, only approximately one third of the grating area is illuminated.

The fourth spectral order of the cross-disperser overlaps with the fifth order in near-IR region. The detector sensitivity in this part of the spectrum was surprisingly high. We installed a shortpass prefilter near the spectrograph slit (item 3 in Figure~\ref{F-optical_layout}) to block the IR part of the spectrum. The prefilter is an ultra wideband optical colored glass filter with a constant transmission within the working wavelength range of the instrument.

\begin{figure}    
\centerline{\includegraphics[width=1.\textwidth,clip=]{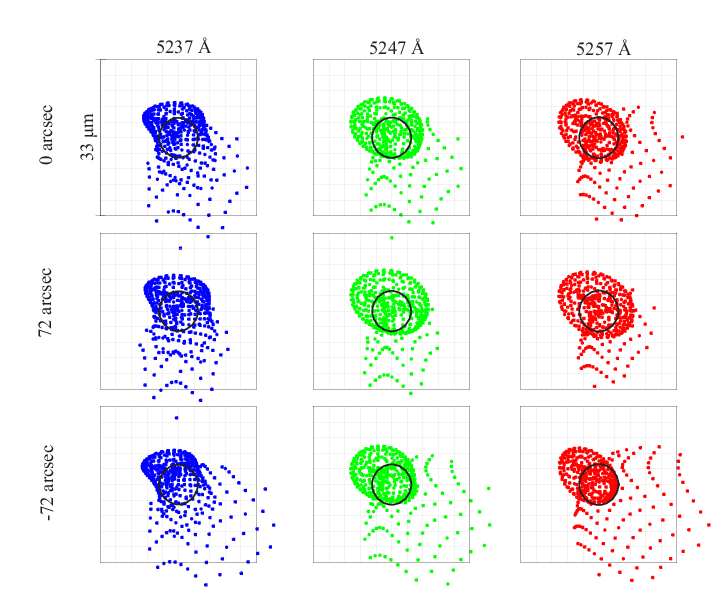}}
\small
        \caption{
        Spot diagrams at the detector sensor plane of the combined telescope and spectropolarimeter optical path for 0, 72, and -72 arcsec field (the angle between the parallel incoming light bundle from the object and the optical axis of the telescope) at three wavelengths (columns) within the 65$^{th}$ spectral order. The size of each large square box is 33$\times$33~$\mu$m$^{2}$, \textit{i.e.} 3$\times$3 detector pixels. The radius of the Airy disk (shown with black circle) is 4.2~$\mu$m.
        }
\label{F-spot_diagram}
\end{figure}

The outgoing beam from the cross-disperser is tilted with respect to the horizontal plane by 39 degrees. The cross-desperser directs the light to the spectrograph camera lens (item 7 in Figure~\ref{F-optical_layout}). The camera is an achromatic doublet with the focal length of 780 mm. The clear aperture of the camera is 155~mm resulting in approximately $f/5$ focal ratio. The camera lens were fabricated at the Crimean Astrophysical Observatory as well from the analogues of fluorite F2 and S-BSL7 glasses with an air gap between lenses. All the surfaces of the lenses are spheres. The camera focal length is constrained by the requirement to image the single element of the spatial resolution, i.e. the width of the slit, to the range of one to three detector pixels. The fulfilment of this requirement can be demonstrated by the spot diagram of the spectrograph at the detector sensor plane, which is shown in Figure~\ref{F-spot_diagram}. The plot shows the spot diagram of combined telescope and spectrograph optical path. The spot diagram is shown for three wavelengths within the spectral order 65 for three fields. The size of the square boxes in the plot is 33 $\mu$m$\times$33~$\mu$m. It can be seen from Figure~\ref{F-spot_diagram} that the entire optical system is close to diffraction limit shown by the 8.4 $\mu$m in diameter Airy disk: most of the energy is concentrated within an area of approximately 16~$\mu$m in diameter. The spectrograph was designed to use a CCD based detector with the pixel size of 13.3$\times$13.3~$\mu$m$^2$. However, we found that particular device was too slow for our goals. The actually installed detector has an sCMOS sensor with 11$\times$11~$\mu$m$^2$ pixel size. Hence, the width of the slit image at the detector is mapped to less than 2 pixels. The detector (item 8 in Figure~\ref{F-optical_layout}) is based on a 2048$\times$2048 pixel GSENSE400BSI sensor by Gpixel and provides more than 90\% quantum efficiency within the wavelength range of SP. With the 200 arcsec field-of-view (see Section~\ref{S-analyzer}), taking into account the collimator to camera focal length ratio $f_{\rm coll}/f_{\rm cam} \approx 7.6$ and solar image scale of 244~$\mu$m~arcsec$^{-1}$ at the telescope focal plane, the height of a spectral order is 500 pixels with the spatial sampling along the slit height of 0.4 arcsec~pixel$^{-1}$. Hence, three spectral orders (64 to 66) are readily captured by the detector simultaneously.

The detector is mounted on a linear motorized translation stage enabling fine focusing of the detector. The liquid cooling of the detector is performed by a thermoelectric chiller located in a separate room to decrease the vibrations within the spectrograph room.

\begin{figure}    
\centerline{\includegraphics[width=1.\textwidth,clip=]{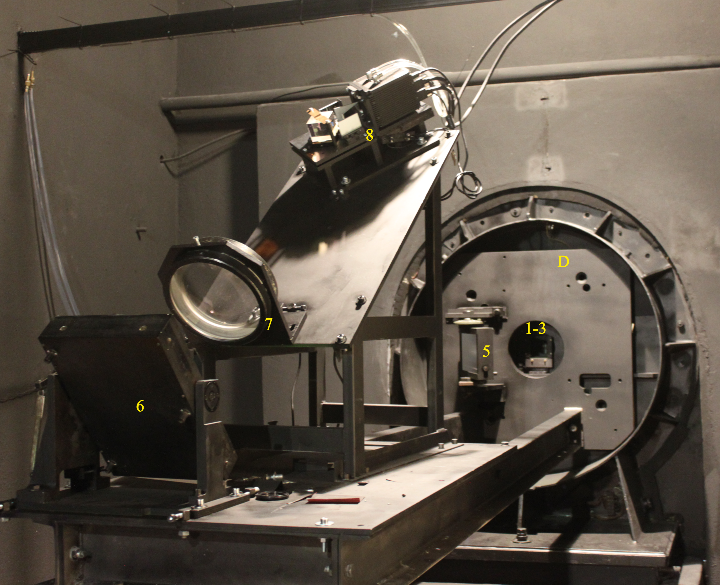}}
        \caption{
        Image of the \textit{Spectro-Polarimeter} with labelled optical elements. The spectrograph optical units (except for the collimator) are mounted on the steel disk separating the spectrograph from the telescope tower (items 1-3, 5) and on a steel optical table (items 6-8). The part of the disk separating the spectrograph is marked with letter `D' in the image. The collimator is installed on a separate tripod (not shown in the image). The prefilter (item 3) masks the spectrograph slit and the polarization modulator. The labelling of the elements is the same as in Figures~\ref{F-optical_layout} and \ref{F-pbs}.
        }
\label{F-spectrograph}
\end{figure}

All the elements of the spectrograph except for the collimator are mounted on the steel disk separating the spectrograph from the telescope tower and on a steel optical table (Figure~\ref{F-spectrograph}). One side of the table is fastened to the steel disk while the second side leans on a steel tripod with a variable height. The collimator is installed on a separate tripod. The optical element mounts are made of either steel or anodized aluminium with black coating. The walls of the spectraograph room are also covered with black mat color to decrease the intensity of scattered light within the spectrograph. The room is not equipped with any kind of ventilation to avoid spectrum degradation by air motions. Instead, low humidity and constant temperature are maintained within the room.

\subsection{Polarization Analyzer}
     \label{S-analyzer}

Polarization analyzer is intended to measure the intensity of polarized light by separating each polarization state or their sum either temporarily or spatially. We refer the reader to \citet{delToroIniesta2003} for a comprehensive review on spectropolarimetry techniques. Our SP is a dual-beam spectropolarimeter with temporal modulation of the polarization. The polarization analyzer of SP is comprised of two units -- the polarization modulator and the polarization beam splitter.

The polarization modulator is a waveplate installed immediately after the slit of the spectrograph. The waveplate is mounted in a motorized rotary stage. The stage provides precise waveplate rotation at the rate of up to three rotations per second. Due to reference and position encoder, the stage ensures excellent repeatability of the waveplate position with the positioning accuracy of 0.1 degree. The stage is controlled via Ethernet by means of a stepper motor controller. 

Depending on the observational objective, we can use either 0.25$\lambda$ or 0.353$\lambda$ achromatic waveplate in the polarization modulator. The former is convenient when the information on circular polarization only is sufficient. The latter provides the highest efficiency of the polarimeter \citep[see][for details on polarimeter efficiency calculations]{delToroIniesta2003}. The clear aperture of both waveplates is 48 mm. The waveplate vignettes the beam transmitted through the 70~mm in height slit and actually defines the field-of-view of the instrument.  

\begin{figure}    
\centerline{
         \includegraphics[width=0.5\textwidth,clip=]{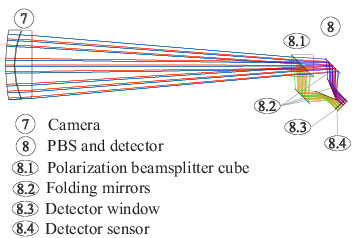}
         \includegraphics[width=0.5\textwidth,clip=]{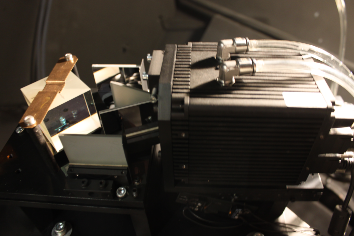}
        }
             
\caption{
            Left -- Ray tracing of the camera and polarization beam splitter. Polarization beamsplitter cube (item 8.1) separates orthogonally polarized beams. Four flat folding mirrors (item 8.2) recombine the beams at the detector sensor plane (item 8.4). The labelling is the same as in Figures~\ref{F-optical_layout} and \ref{F-spectrograph}. Right -- The image of the polarization beam splitter and detector (item 8 in the left panel).
        }
\label{F-pbs}
\end{figure}

The polarization beam splitter is made of the stock available cube with the linear polarizer cemented in the cube diagonal. The cube creates two orthogonally polarized beams separated by 90 degrees. The beams are then combined on the detector by four flat folding mirrors (Figure~\ref{F-pbs}). The beam splitter is installed just ahead of the detector to ensure the longest common path of both orthogonally polarized beams through the spectrograph.

The extinction ratio of the transmitted and reflected by the polarization beam splitter polarized light reaches 1000:1. However, this contrast decreases if the incidence angles exceed certain value. The angular acceptance range in our design is 4.5 degree that is more than two times higher than the optimal value. As a consequence, we expect lower contrast and lower analyzer efficiency. Our calculations revealed that, with the current camera, the addition of relay optics similar to ViSP analyzer \citep[Figure~5 in][]{deWijn2022} would not decrease the incident angles significantly. On the other hand, this decision would result in considerable spectral resolution loss.

\begin{figure}    
\centerline{\includegraphics[width=1.\textwidth,clip=]{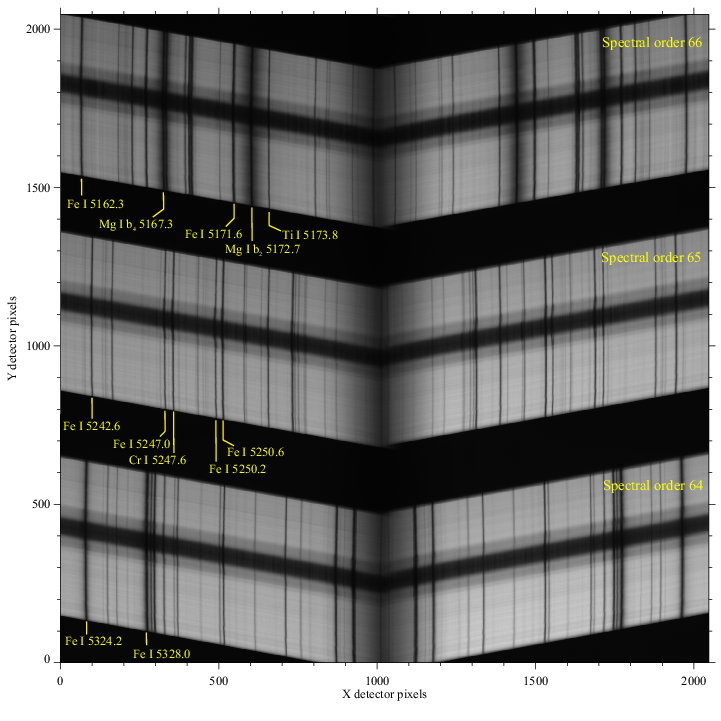}}
        \caption{
        A typical raw single-exposure spectrum taken by SP. Three spectral orders (64 to 66) in two orthogonally polarized states are mapped symmetrically with respect to the central columns of the detector. Spectral lines are roughly co-aligned with the columns along detector $Y$ direction while the spectral dispersion is in the direction of tilted orders with longer wavelengths being closer to the detector center. Several spectral lines are identified in each spectral order. The slit of SP was set across a sunspot. The displacement of spectral line centers in penumbra and umbra due to strong magnetic fields is clearly seen.
        }
\label{F-spectrum_raw}
\end{figure}

The size of the polarization beam splitter face is 50$\times$50 mm$^2$. The entire unit is placed close to the detector entrance window to prevent vignetting (see right panel in Figure~\ref{F-pbs}). The cube and mirrors mount provide precise 3D adjustment of each element separately to achieve the best and simultaneous focusing of both images at the detector sensor plane. The final spectrum in two polarization states recorded by the detector is shown in Figure~\ref{F-spectrum_raw}.

\section{Spectra Reduction}
\label{s-reduction}

The first tests of SP in spectropolarimetric mode were carried out between April and July 2023. A typical raw spectrum acquired by the instrument is shown in Figure~\ref{F-spectrum_raw}. Three spectral orders in two orthogonally polarized states are mapped symmetrically with respect to the central columns of the detector. Spectral lines are roughly co-aligned with the columns along detector $Y$ direction while the spectral dispersion is in the direction of tilted orders with longer wavelengths being closer to the detector center. Several spectral lines are identified in Figure~\ref{F-spectrum_raw}. Spectral order 64 covers the Fe~I 5324.2~\AA\ spectral line \citep{Ai1982} that is traditionally used to probe the solar photospheric magnetic field by Chinese scientific facilities \citep[\textit{e.g.}][]{Xu2024}. The slit of SP in Figure~\ref{F-spectrum_raw} was set across a sunspot. The displacement of spectral line centers in penumbra and umbra due to strong magnetic fields is clearly seen.

The raw spectrum recorded by the detector (Figure~\ref{F-spectrum_raw}) is a typical multi-order echelle spectrum usually acquired for point sources during night time observations. The reduction of the spectra is a rather complicated and a non-trivial task \citep[\textit{e.g.,}][]{Piskunov2002}. The intrinsic properties of echelle spectra are curvatures of both spectral orders and spectral lines \citep[\textit{e.g.}, section 5.8 in][]{Eversberg2015}. In contrast to point-source night-time observations, during the solar spectropolarimetry we have to keep the spectral information along the slit direction. That is why numerous packages for echelle spectra reduction \citep[\textit{e.g.,}][]{Piskunov2021, Galazutdinov2022} are not directly applicable in our case. Nevertheless, we use certain approaches from those packages.

In addition to spectra and order curvatures, the polarization beam-splitter distorts each of the beams in its own way due to slightly different mirror inclinations, non ideal parallelism of the beamsplitter cube, \textit{etc}. Intensities of the orthogonally polarized beams are also different due to non-equal transmission or reflection of polarized light by each optical element of the spectrograph. Dual-beam spectropolarimetry implies merging of two orthogonally-polarized beams to derive Stokes vector components. This procedure decreases the cross-talk between the Stokes components due to residual image motion caused by atmospheric seeing \citep[\textit{e.g.}][]{Lites1987, Judge2004}. Hence, two images of orthogonally polarized spectra must be precisely co-aligned before merging. To do this, in our data reduction routine we primarily adopt approaches developed for data reduction of solar dual-beam spectropolarimeters outputs, namely for the \textit{Hinode Spectro-Polarimeter} \citep{Lites2013}, the \textit{Advanced Stokes Polarimeter} \citep{Skumanich1997}, and for the \textit{Polarimetric Littrow Spectrograph} \citep{Beck2005}.

Our procedure of the spectra reduction includes subtraction of the dark bias and of the scattered light from each frame, correction for the inhomogeneous slit width and different detector pixel sensitivity, normalization to a unity continuum intensity, correction for different dispersion in each polarization, merging of two orthogonally polarized spectra, elimination of the spectral order and spectral line curvatures. These steps mostly rely on the functions describing the properties of the spectropolarimeter, such as the dark bias level, slit intensity variations, spectral curvature, pixel-by-pixel gain, \textit{etc}. Most of the functions are derived from a series of several thousands of calibration frames taken near the center of the solar disk during each set of observations. While capturing the calibration frames, the solar image at the spectrograph entrance is continuously shifted in different directions to smooth away the intensity variations related to real features at the solar surface and limb darkening. Below we briefly describe the routine of data reduction. A more comprehensive paper on the data reduction technique and on the polarization calibration of the instrument is in preparation.

At the first step a dark image is subtracted from each recorded frame. To acquire the dark image, we cover the spectrograph slit during the observation set and record several hundreds of frames. These frames are averaged to obtain the final dark image. At the next step a scattered light image is subtracted from each raw spectrum. The procedure of scattered light image subtraction is similar to that used in the DECH package for echelle spectra reduction \citep{Galazutdinov2022}. For each column of the recorded spectrum (slices along the slit height in $Y$ direction in Figure~\ref{F-spectrum_raw}) we select low intensity pixels belonging to inter-order minima. The values in the pixels are approximated by a second order polynomial. A 2048$\times$2048 pixels scattered light frame is constructed by filling the columns with the values calculated using the derived polynomials. The polynomials differ slightly from column to column (in $X$ direction in Figure~\ref{F-spectrum_raw}) due to noise in the data. To get a smooth image, the constructed scattered light frame is then approximated with a second order surface. The surface is considered as a final scattered light image and is calculated for each spectrum individually.

The following reduction steps are applied for each spectral order and for each polarization state separately. The reduction is performed for the 800-pixel wide left-hand and right-hand parts of the raw spectrum to exclude the central columns with reduced intensity and spectral orders overlapping ($X=[800:1200]$ in Figure~\ref{F-spectrum_raw}). We utilize the calibration frames to retrieve the function of the slit width variation $\Delta W_{L,R} (y)$, where $y$ is the distance along the slit and the indices $L$ and $R$ in function notations stand for left and right parts of the spectrum image in Figure~\ref{F-spectrum_raw}, \textit{i.e.} for different polarization states. To retrieve $\Delta W_{L,R} (y)$, we select a central vertical slice of the spectral order as a reference one. Then we perform a cross-calibration of the reference slice with each slice of the spectral order and derive the relative shift of the order for each $X$ coordinate. To ensure the best sub-pixel accuracy, the cross-calibration is performed for numerical second derivative of the slices \citep{Lites2013}. The shift values are fitted with a second-order polynomial to derive the function of the spectral order curvature $\delta_{L,R}(x)$. Using $\delta_{L,R}(x)$, we again calculate the shift of each vertical slice of the spectrum order and shift the slice with a sub-pixel accuracy, thus eliminating the order curvature. Then the intensities within each row (along the dispersion) are averaged resulting in the intensity variations function along the slit for a particular calibration frame. The same procedure is repeated for every calibration frame. The retrieved functions are co-aligned by a cross-correlation technique and averaged. Since the calibration frames are taken at random positions at the solar surface, the averaging eliminates the intensity variations due to solar features and limb darkening. The average is further divided by its median value yielding $\Delta W_{L,R} (y)$.

The acquired function $\Delta W_{L,R} (y)$ is utilized to correct the intensity variations due to inhomogeneous slit width in the data frames. Again, similar to calibration frames, we derive the spectral order curvature, calculate the spectral order shift for the current slice with respect to some reference slice, and shift $\Delta W_{L,R} (y)$ by this value along the slit direction \citep[see Section~2.9 in][]{Lites2013}. Then the slice is divided by $\Delta W_{L,R} (y)$ and the procedure is repeated for the next slice.

The pixel-by-pixel gain variation function $G_{L,R}(x,y)$ is obtained by imaging the spectrum of an incandescent lamp with a flat rectangular glower. Large scale variations of the spectrum are corrected in the same manner as described above for the solar spectra. The solar spectra are divided by the normalized to unity continuum $G_{L,R}(x,y)$. By this point, no corrections for the spectral order or spectral line curvatures are made.

To correct the spectral order curvature, the function $\delta_{L,R}(x)$ is calculated for each spectrum. Due to wedging of the waveplate, spectrum images at the detector exhibit slight displacement by several pixels as the polarization analyzer rotates. Hence, the spectral order curvature is derived for each spectrum individually. The spectral image columns are shifted according to the value of $\delta_{L,R}(x)$ with a sub-pixel accuracy. As a result, we get a rectangular shaped spectral image. 

We further perform normalization of the spectrum to a unity continuum. Reduced spectra for a single set of observations are averaged yielding an average flat image. For this image we further average all the columns to obtain a master one-dimensional spectrum similar to that shown in Figure~\ref{F-spectrum_slices}. In this spectrum we manually select the points of the continuum regions outside spectral lines. These points are approximated by a third-order polynomial $C_{L,R}(x)$ that is used as a normalization function. By dividing columns of each spectrum by $C_{L,R}(x)$, we eliminate large scale variations of the intensity in the dispersion direction and scale the spectrum to a unity continuum.

The next step is to scale spectra in different polarization states to an equal linear dispersion. For this purpose we use the average of the reduced calibration spectra. Again, since calibration spectra are taken at different positions at the solar surface, its average is free from Zeeman patterns of magnetic features and systematic Doppler shifts due to solar rotation. The shifts between the same spectral lines in the ``left'' and ``right'' average spectra are derived by a cross-correlation technique. The dependence between the shifts and the line positions in the spectra is fitted by a linear function. The slope of the function provides the ratio between the linear dispersion in the ``left'' and ``right'' polarization states. The wavelength of the spectra from the right part of the detector is interpolated to match that of the ``left'' spectra.

The curvature of the spectral lines is also determined from the averaged calibration spectra. For some interval around a spectral line, we select a row (\textit{i.e.} in the dispersion direction) from the middle of the spectrum image as a reference spectral line profile. Then, for each row of the spectra we derive the shift between the current and reference profiles by a cross-correlation technique. The resulting shifts for each row are fitted by a third-order polynomial to get the spectral line curvature function $\delta \lambda (y)$. The difference of the spectral line wavelength position near the center and at the spectral order edges reaches 4 detector pixels. Function $\delta \lambda (y)$ is used to eliminate the spectral line curvature in the spectra by shifting the image rows.

Finally, reduced spectra for orthogonal polarization states are merged. All the previous steps aligned the spectra both along the slit and in the dispersion direction. The same intensity of the images is ensured by the normalization to unity continuum. The spectral images are summed to provide Stokes $I$ and subtracted to derive Stokes $Q$, $U$, and $V$ components.

\section{Polarization Calibration}
\label{S-polarization}

The primary goal of SP is to measure the polarization state or the Stokes vector $S_{Sun}=[I, Q, U, V]^T$ of the incoming solar light. The telescope and spectropolarimeter change the polarization state of the light by introducing a cross-talk between the Stokes components. Using the Mueller matrices formalism \citep[\textit{e.g.}][]{delToroIniesta2003}, the measured Stokes vector can be represented as $S_{obs}= \mathbf{M_{SP}} \mathbf{M_{tel}} S_{Sun}$, where $\mathbf{M_{SP}}$ is the 4$\times$4 response matrix of SP and $\mathbf{M_{tel}}$ is the matrix describing the polarization changes by the telescope optics. During polarization calibration we measure $\mathbf{M_{SP}}$ and use an approach described by \citet{Jaeggli2022} to derive $\mathbf{M_{tel}}$.

STT/SP uses a standard polarization analyzer consisting of a waveplate rotator and a linear polarizer to decode the polarization state of $S_{obs}$. The Mueller matrix of an ideal waveplate $M_{WP}$ is
\begin{equation}
	M_{WP}(\theta, \delta) = 
	\begin{bmatrix}
	1       & 0 & 0 & 0 \\
	0       & \cos^2 2\theta + \sin^2 2\theta \cos\delta & \cos 2\theta \sin 2\theta (1 - \cos \delta )& \sin 2\theta  \sin \delta \\
	0       & \cos 2\theta \sin 2\theta (1 - \cos \delta)& \cos^2 2\theta \cos \delta + \sin^2 2\theta  & -\cos 2\theta \sin \delta \\
	0       & -\sin 2\theta \sin \delta & \cos 2\theta \sin \delta & \cos \delta 
	\end{bmatrix},
\end{equation}
where $\delta$ is the waveplate retardance and $\theta$ is the angle between the fast axis of the waveplate and some reference direction. For a linear polarizer, one can write
\begin{equation}
	M_{LP}(\beta) = 
	\begin{bmatrix}
		1       & \cos 2 \beta & \sin 2 \beta & 0 \\
		\cos 2 \beta & \cos^2 2\beta & \cos 2\beta \sin 2\beta & 0 \\
		\sin 2 \beta & \cos 2\beta \sin 2\beta & \sin^2 2\beta  & 0 \\
		0       & 0 & 0 & 0 
	\end{bmatrix},
\end{equation}
where $\beta$ is the angle of the polarizer rotation. In SP, the polarization beamsplitter acts as two linear polarizers rotated by 90$^{\circ}$ with the reference direction coinciding with the direction of one of the polarizer. By multiplying the elements in the ideal SP response matrix $\mathbf{M_{LP}}(0^{\circ}, 90^{\circ}) \mathbf{M_{WP}}(\theta, \delta)$ and taking into account that the detector is sensitive only to the Stokes $I$ component of the beam, we obtain the following expression for the detector signal 
\begin{equation}
\begin{split}
	D^{L,R}(\theta, \delta) =  \frac{r^{L,R}}{2} [\ I \pm Q (\cos^2 2 \theta + \sin^2 2 \theta \cos \delta) \\
	\pm U \sin 2 \theta \cos 2 \theta (1 - \cos \delta) \\  \mp V \sin 2 \theta \sin \delta ],
\end{split}
\end{equation}
where $I$, $Q$, $U$, and $V$ are the Stokes components of the incoming beam \citep{Landi2004, Beck2005}, indices $L$ and $R$ stand for different sides of the detector and correspond to plus and minus signs in the expression. The coefficients $r^{L,R}$ describe the throughput of SP for ``left'' and ``right'' polarization states and are equal to unity due to normalization at the spectra reduction stage. For a quarter waveplate retarder ($\delta$=90$^{\circ}$), it is sufficient to perform four measurements taken with 22.5$^{\circ}$ modulator step to retrieve full Stokes vector:
\begin{equation}
	\label{eq_ideal}
\begin{split}
	D^{L,R}(0^{\circ}, 90^{\circ}) = \frac{1}{2}[ I \pm Q ];\\
	D^{L,R}(22.5^{\circ}, 90^{\circ}) = \frac{1}{2}[ I \pm (0.5 Q + 0.5 U + 0.7V) ; \\
	D^{L,R}(45^{\circ}, 90^{\circ}) = \frac{1}{2}[ I \pm (V) ]; \\
	D^{L,R}(67.5^{\circ}, 90^{\circ}) = \frac{1}{2}[ I \pm (0.5 Q - 0.5 U + 0.7V).
\end{split}
\end{equation}

In practice the SP response matrix deviates from an ideal one. We use three calibration units to derive the actual response matrix. The calibration units are linear, right-hand, and left-hand circular polarizers installed in front of the spectrograph slit. The polarizers can be rotated with a high precision. The light source for the polarization calibration is the same incandescent lamp used to obtain the pixel-by-pixel gain variation function. The light of the lamp is assumed to be unpolarized, \textit{i.e.} the Stokes vector of the lamp is $S_{\rm lamp} = [1, 0,0,0]^T$. The calibration procedure consists of the measurements of the Stokes vector $S_{\rm calib}=\mathbf{M_{CU}} S_{\rm lamp}$ created by the calibration units for a half rotation with 10$^{\circ}$ step, where $\mathbf{M_{CU}}$ is a precisely measured response matrix of either linear, or left-hand, or right-hand circular polarizer. The actually measured Stokes vector $S_{\rm obs}$ (using Equation~\ref{eq_ideal}) equals $S_{\rm obs} = \mathbf{M_{SP}} S_{\rm calib}$ with 15 unknown elements in $\mathbf{M_{SP}}$ (the element $m_{00}$ equals unity after normalization). $\mathbf{M_{SP}}$ is derived from a calibration data set of 18 (the number of steps) by 3 (the number of calibration units) linear equation systems \citep[\text{see, e.g., appendix in}][]{Beck2005}.

The determination of the telescope response matrix $\mathbf{M_{tel}}$ is more complicated. Each mirror of the telescope changes the polarization of the reflected beam to a certain extent depending on the angle of incidence and the properties of the mirror coating. Although the primary and secondary mirrors are stationary, the main issue is the coelostat mirror that rotates permanently resulting in continuous changes of the incidence angle. The one way is to apply a similar approach with a rotating calibration polarizer or waveplate for the telescope. However, the installation of a rotating waveplate in front of a 1.2 m coelostat mirror is beyond our technical capabilities. The second way is to develop the polarization model of the telescope \citep[\textit{e.g.}][]{Skumanich1997, Beck2005b}. However, the model would definitely provide low accuracy due to uncertainties in the telescope mirror parameters. So far we decided to apply an approach proposed recently by \citet{Jaeggli2022} for \textit{ad hoc} correction of the derived Stokes vector. The approach relies on the supposition that the Stokes components are either symmetrical ($I$, $Q$, $U$) or anti-symmetrical ($V$) with respect to spectral line center. The response matrix of any optical non-depolarizing system can be represented as a multiplication of matrices of a general elliptical retarder and a general elliptical diattenuator. Each of the matrices can be described by three variables \citep[see Section~2 in][]{Jaeggli2022}. Using the observational data, the diattenuator matrix can be derived by minimizing the correlation between uniformly signed Stokes $I$ and alternating Stokes $Q$, $U$, and $V$. The elliptical retarder matrix is found by minimization of the correlation of $Q$ to $V$ and $U$ to $V$ \citep[Equations 16 and 17 in][]{Jaeggli2022}. The authors tested the approach on contaminated \textit{Hinode}/SOT-SP data and obtained better than $10^{-3}$ polarimetric accuracy in the recovered data. Our tests revealed that the polarimetric accuracy achieved by our instrument is of about $5 \times 10^{-3}$ or better and is limited mostly by high noise of the detector. According to \citet{Ichimoto2008}, the polarization accuracy better than $10^{-3}$ is hardly available during high and moderate resolution ground-based spectropolarimetry due to seeing. In our case, the jitter of the solar image at the slit makes it impossible to increase the integration time and, as a consequence, to decrease the noise in the data substantially. We hope that our further efforts on image stabilization will help us to get over this issue.

\section{Modes of Observations}
     \label{S-modes} 

The SP instrument provides a great flexibility in the observational modes. In general, the data collecting assumes scanning of the solar surface image perpendicular to the spectrograph slit. The scanning is synchronized in some way with the rotation of the waveplate in the polarization analyzer, which modulates the intensity of the polarized light. The intensity variations of the recorded spectra are then converted to the spectral profiles of the Stokes vector. The details of this algorithm relies on the particular scientific task. Below we briefly describe the modes of operation provided by the instrument.

The most common mode is the full Stokes vector polarimetry. In this mode, the variation of the data collecting algorithm is available. First, the image of the Sun is kept stationary at the slit while the polarization modulator rotates by 22.5 degrees in each step. Then the data are recorded by the detector in each modulator position. Half turn of the modulator is sufficient to get information on the full Stokes vector. With the maximum frame rate of the detector of 24 frames~s$^{-1}$ and the typical single exposure of 10 ms, it takes less than 0.5 s to take 8 raw spectrum frames. These 8 frames are processed later to derive 4 unknown Stokes vector components by solving an overdetermined system of 8 linear equation. Approximately 1.5 s more is needed to move the solar image to the next slit position. Hence, a single slit position is sampled in 2 s. The entire map of the solar region of interest is combined from the acquired solar slices. Using the slit width of 1 arcsec, it takes approximately 400 s to observe a 170$\times$200 arcsec$^2$ region of the Sun. The integration time can be increased to achieve higher signal to noise ratio required for reliable polarimetric measurements within dark faint regions of sunspot umbra. Another option is to increase the duty cycle by continuous recording of the spectrum with moving slit and spinning modulator. In such a case, the solar image shift during a single modulator rotation must be less than the half of the slit width.

The second polarimetric mode is continuous movement of the solar image across the slit while keeping the polarization modulator at a constant position. Once scanned at a certain polarization state, the polarization modulator rotates by a certain angle and the scan of the region of interest is repeated. The weak point of this approach is that different polarization states cannot be actually recorded for exactly the same region of the Sun. This issue can be neglected to some degree if the oversampling occurs, \textit{i.e.} the step of the slit is sufficiently lower than the spatial resolution of the telescope. This mode provides several times faster data acquisition due to higher duty cycle. Each scan is used to create a data cube (spatial 2D map and 1D spectrum) of the solar region at a certain polarization state. The Stokes vector components then derived by combining these cubes. The best practice to carry out full Stokes vector spectrpolarimetry at STT/SP is yet to be elaborated.

The next mode is a fast intensity spectroscopy although providing some reduced information on the polarization state of the light. In this mode the solar image is moving at a constant rate across the slit. The polarization modulator waveplate can be removed or set to encode a certain polarization state, say, to extract the circular polarization only. The spectrum intensity can be roughly derived by summing two spectra acquired by the detector in two polarization states. In addition to intensity, information on circular polarization can be easily obtained by subtracting the spectra. The precise intensity measurement not contaminated by polarization is really not possible: every optical element of the spectrograph varies the orthogonally polarized components of the light in a different way (for instance, the light reflection by the echelle grating along and across the grooves). As a result, the polarization beam splitter would change the intensity of the light at the detector to some extent. However, this mode can be used in the observations where a high temporal cadence is required, for example, in the analysis of wave propagation. The cadence can be as high as 30 s when observing a moderate-size active region.

Both spectrpolarimetric and fast spectroscopic mode can be used with a stationary solar image at the slit (sit-and-stare). This mode is required to capture very fast processes in, say, solar flares at cadences of seconds. In this mode, only the solar spectrum along the slit is available.


\section{Preliminary Results}
     \label{S-results} 

\begin{figure}    
\centerline{\includegraphics[width=1.\textwidth,clip=]{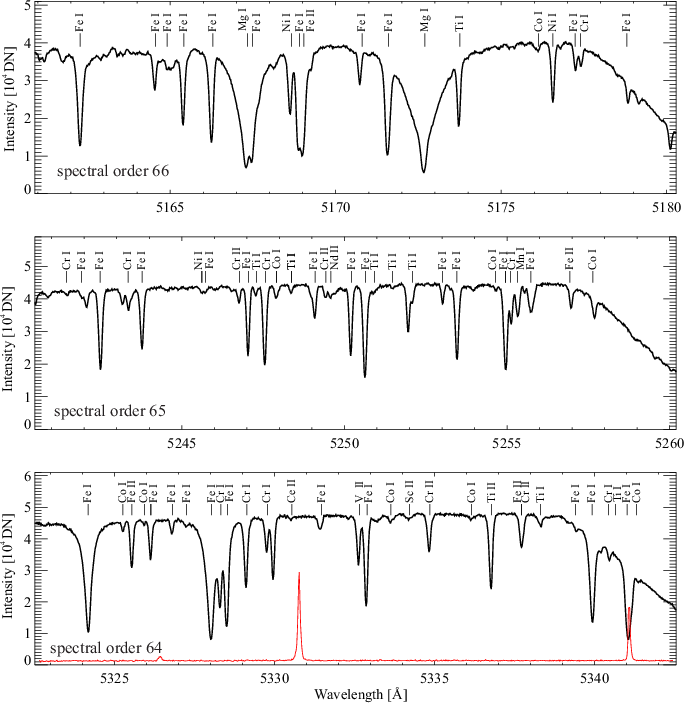}}
\small
        \caption{
        Slices of the raw spectrum (Figure~\ref{F-spectrum_raw}) along the spectral dispersion for spectral orders 64 (bottom), 65 (middle), and 66 (top). The signal-to-noise ratio of the spectrum exceeds 200. The line identification is from \citet{AllendePrieto1998}. The red curve in the bottom panel demonstrates a reference neon spectrum.
        }
\label{F-spectrum_slices}
\end{figure}

\begin{figure}    
\centerline{\includegraphics[width=.45\textwidth,clip=]{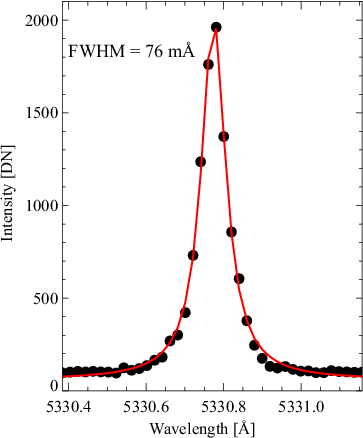}}
\small
        \caption{
        The spectral line of a reference neon lamp recorded by SP (black circles). Red line shows the best approximation of the data points by a Lorentz spectral profile. The FWHM of the profile is 76 m\AA.
        }
\label{F-neon}
\end{figure}

Figure~\ref{F-spectrum_slices} demonstrates the slices of the quiet Sun raw spectrum (Figure~\ref{F-spectrum_raw}) along the spectral dispersion. Each spectral order is more than 15~\AA\ wide. The most prominent spectral lines are identified in the plots using the solar spectrum data by \citet{AllendePrieto1998}: the instrument observes dozens of magnetically sensitive spectral lines simultaneously. The spectrum is recorded with 10 ms exposure to prevent saturation. The signal-to-noise ratio of this single exposure exceeds 200. The red line in the lower panel of Figure~\ref{F-spectrum_slices} overplots the spectrum of a reference neon lamp. The neon spectral line around 5330.8~\AA\ is shown in more detail in Figure~\ref{F-neon}. The FWHM of the profile is 76~m\AA. Assuming that the neon linewidth is determined by the SP instrumental function, we can estimate the spectral resolution to be of about $R \approx 70,000$.


\begin{figure}    
\centerline{\includegraphics[width=1.\textwidth,clip=]{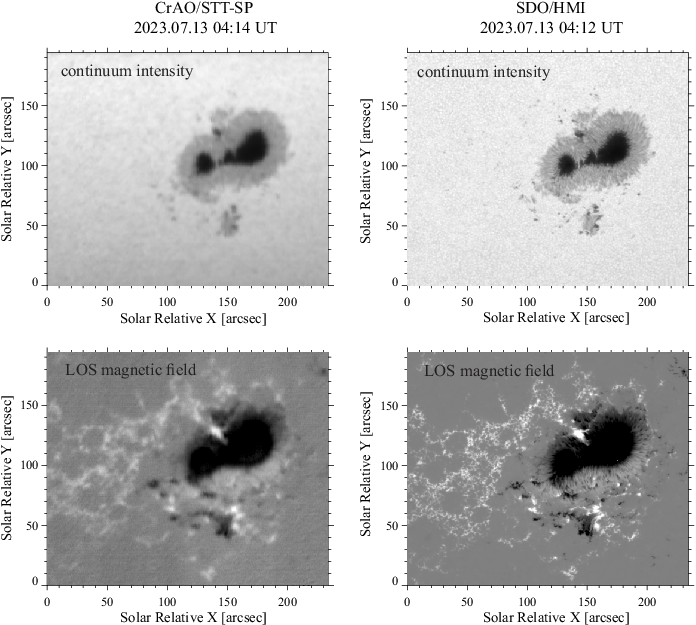}}
\small
        \caption{
        Co-temporary continuum intensity images (top panels) and longitudinal magnetic field maps (bottom panels) of NOAA AR 13363 acquired on 2023.07.13 by CrAO/STT-SP (left panels) and SDO/HMI (right panels). The CrAO/STT-SP magnetic field map is obtained using Fe~I 5250.2~\AA\ spectral line. The images are rotated by 4 degrees counter-clockwise with respect to the solar North-South direction. The magnetograms are scaled from -1000~Mx~cm$^{-2}$ (black) to 1000~Mx~cm$^{-2}$ (white). The field-of-view is 235$\times$195 arcsec$^{2}$. 
        }
\label{F-bz}
\end{figure}

The CrAO/STT-SP maps of the continuum intensity and of longitudinal magnetic field of NOAA AR 13363 are shown in Figure~\ref{F-bz}. The raster scans were obtained on 2023 July 13 near 04:12 UT. The active region was observed in fast spectroscopy mode using $\lambda/4$ waveplate with the fast axis rotated by 45$^{\circ}$ with respect to the fast axis of the polarization beam splitter. Hence, with no regard to the instrumental polarization, the detector recorded Stokes $\frac{1}{2}(I+V)$ and $\frac{1}{2}(I-V)$ signals at its left and right sides according to the third expression in equations~\ref{eq_ideal}. The data were sampled in 400 slit positions with the slit step of about 0.59 arcsec and 10 ms integration time per position. The total resulting field-of-view was 240$\times$200~arcsec$^2$. Fe~I 5250~\AA\ spectral line was used to calculate longitudinal magnetic field by the center-of-gravity approach \citep{Semel1967}. The method treats the difference between the center-of-gravity of Stokes $I+V$ and $I-V$ profiles as a proxy of longitudinal magnetic field. The obtained maps are consistent with that provided by the \textit{Helioseismic and Magnetic Imager} \citep[HMI,][]{Schou2012} on board the \textit{Solar Dynamics Observatory} \citep[SDO:][]{Pesnell2012} shown in right panels of Figure~\ref{F-bz}. Table~\ref{T-summary} presents the summary of the measured instrument parameters.

\begin{table}
\caption{ Summary of the measured (as-built) parameters of SP
}
\label{T-summary}
\begin{tabular}{lll}     
\hline                     
Parameter & Value & Notes \\
  \hline
Wavelength ranges &5161 to 5180~\AA &spectral order 66\\
                &5241 to 5260~\AA &spectral order 65\\
                &5323 to 5340~\AA&spectral order 64\\
Spectral resolution & $\ge 70,000$ & \\
Spectral sampling & 20~m\AA~pixel$^{-1}$ &\\
Spatial sampling & 0.4~arcsec~pixel$^{-1}$ &along the slit\\
Field-of-view & 200~arcsec &along the slit \\
Cadence & $\le$ 10 min & spectropolarimetric mode\\
 & $\le$ 1 min & intensity mode\\
Polarimetric accuracy & $\le 5 \times 10^{-3}$ & \\
Context imager FOV&150$\times$150~arcsec$^{2}$ &in TiO 7057~\AA\ band\\
\hline
\end{tabular}
\end{table}

\section{Conclusions}
     \label{S-conclusions}

SP is a new instrument intended for strictly simultaneous spectropolarimetric observations of the Sun within three spectral ranges. The tool provides new possibilities in exploration of the magnetic field and plasma thermodynamics in the photosphere and lower chromosphere. A great flexibility in the available operation modes will satisfy most of the scientific cases requirements regarding the temporal resolution, signal-to-noise ratio, polarimetric accuracy, \textit{etc}.

Our plans include further upgrades of the instrument. We intend the implementing of a zero-order adaptive optics by replacing the stationary folding mirror directing the solar beam to the spectrograph slit by a tip-tilt mirror. We hope this assembly will allow us to improve the spatial resolution of the derived solar maps making the instrument even more profitable for the solar community.

\begin{acks}
We are grateful to the anonymous referee whose comments motivated us to improve the instrument calibration and data reduction routine. SDO data courtesy of NASA/SDO and the HMI science team.
\end{acks}

\begin{authorcontribution}
Funding acquisition: V.~Abramenko; Instrument Design: V.~Terebizh; A.~Kutsenko; Instrument Implementation: A.~Dolgopolov, A.~Kutsenko, D.~Semyonov, V.~Skiruta, V.~Lopukhin; Telescope Upgrade: A.~Kutsenko, A.~Dolgopolov, D.~Semyonov, V.~Lopukhin; Software Development: A.~Kutsenko; Data Analysis: A.~Kutsenko, A.~Plotnikov; Writing -- original draft preparation: A.~Kutsenko; Writing - review and editing: all authors.
\end{authorcontribution}

\begin{fundinginformation}
\textit{Andrei B. Severny Solar Tower Telescope} is a scientific facility of the \textit{Crimean Astrophysical Observatory} (CrAO). CrAO is managed by the Ministry of Science and Higher Education of the Russian Federation.
\end{fundinginformation}

\begin{dataavailability}
The data sets collected by STT-SP and analyzed in this work were taken as a preliminary test of the instrument performance during the commissioning phase and are available from the corresponding author on request. The SDO/HMI data are available via JSOC at https://jsoc.stanford.edu/.
\end{dataavailability}



\begin{ethics}
\begin{conflict}
The authors declare that they have no conflicts of interest.
\end{conflict}
\end{ethics}

\bibliographystyle{spr-mp-sola}
\bibliography{spectropol}  

\begin{thebibliography}{45}
\ifx\bisbn     \undefined \def\bisbn  #1{ISBN #1}\fi
\ifx\binits    \undefined \def\binits#1{#1}\fi
\ifx\bauthor   \undefined \def\bauthor#1{#1}\fi
\ifx\batitle   \undefined \def\batitle#1{#1}\fi
\ifx\bjtitle   \undefined \def\bjtitle#1{\textit{#1}}\fi
\ifx\bvolume   \undefined \def\bvolume#1{\textbf{#1}}\fi
\ifx\byear     \undefined \def\byear#1{#1}\fi
\ifx\bissue    \undefined \def\bissue#1{#1}\fi
\ifx\bfpage    \undefined \def\bfpage#1{#1}\fi
\ifx\blpage    \undefined \def\blpage #1{#1}\fi
\ifx\burl      \undefined \def\burl#1{\href{#1}{\textsf{URL}}}\fi
\ifx\href      \undefined \def\href#1#2{#2}\fi
\ifx\betal     \undefined \def\betal{et al.}\fi
\ifx\bctitle   \undefined \def\bctitle#1{#1}\fi
\ifx\beditor   \undefined \def\beditor#1{#1}\fi
\ifx\bbtitle   \undefined \def\bbtitle#1{\textit{#1}}\fi
\ifx\bedition  \undefined \def\bedition#1{#1}\fi
\ifx\bseriesno \undefined \def\bseriesno#1{\textbf{#1}}\fi
\ifx\blocation \undefined \def\blocation#1{#1}\fi
\ifx\bsertitle \undefined \def\bsertitle#1{\textit{#1}}\fi
\ifx\bsnm      \undefined \def\bsnm#1{#1}\fi
\ifx\bsuffix   \undefined \def\bsuffix#1{#1}\fi
\ifx\bparticle \undefined \def\bparticle#1{#1}\fi
\ifx\barticle  \undefined \def\barticle#1{}\fi
\ifx\binstitute  \undefined \def\binstitute#1{#1}\fi
\ifx\bpublisher  \undefined \def\bpublisher#1{#1}\fi
\ifx\doiurl    \undefined \def\doiurl#1{\href{#1}{DOI}}\fi
\makeatletter
\def\safeHref#1#2#3{\in@{http}{#2}\ifin@\href{#2}{#3}\else\href{#1#2}{#3}\fi}
\makeatother
\ifx\adsurl    \undefined
  \def\adsurl#1{\safeHref{https://ui.adsabs.harvard.edu/abs/}{#1}{ADS}}\fi
\ifx\arxivurl  \undefined
  \def\arxivurl#1{\safeHref{http://arxiv.org/abs/}{#1}{arXiv}}\fi
\ifx\botherref \undefined \def\botherref#1{}\fi
\ifx\url       \undefined \def\url#1{#1}\fi
\ifx\bchapter  \undefined \def\bchapter#1{}\fi
\ifx\bbook     \undefined \def\bbook#1{}\fi
\ifx\bcomment  \undefined \def\bcomment#1{#1}\fi
\ifx\oauthor   \undefined \def\oauthor#1{#1}\fi
\ifx\citeauthoryear \undefined\def \citeauthoryear#1{#1}\fi
\def\endbibitem {}
\ifx\bconflocation  \undefined \def\bconflocation#1{#1} \fi

\bibitem[\protect\citeauthoryear{{Ai}, {Li}, and {Zhang}}{1982}]{Ai1982}
\begin{barticle}
\bauthor{\bsnm{{Ai}}, \binits{G.-X.}},
\bauthor{\bsnm{{Li}}, \binits{W.}},
\bauthor{\bsnm{{Zhang}}, \binits{H.-Q.}}:
\byear{1982},
\batitle{{FeI lambda 5324.19 A line forms in the solar magnetic field and the
  theoretical calibration of the solar magnetic field telescope}}.
\bjtitle{Acta Astron. Sin.}
\bvolume{23},
\bfpage{39}.
\adsurl{1982AcASn..23...39A}.
\end{barticle}
\endbibitem

\bibitem[\protect\citeauthoryear{{Allende Prieto} and {Garcia
  Lopez}}{1998}]{AllendePrieto1998}
\begin{barticle}
\bauthor{\bsnm{{Allende Prieto}}, \binits{C.}},
\bauthor{\bsnm{{Garcia Lopez}}, \binits{R.J.}}:
\byear{1998},
\batitle{{A catalogue of accurate wavelengths in the optical spectrum of the
  Sun}}.
\bjtitle{\aaps}
\bvolume{131},
\bfpage{431}.
\doiurl{https://doi.org/10.1051/aas:1998280}.
\adsurl{1998A&AS..131..431A}.
\end{barticle}
\endbibitem

\bibitem[\protect\citeauthoryear{{Babcock}}{1953}]{Babcock1953}
\begin{barticle}
\bauthor{\bsnm{{Babcock}}, \binits{H.W.}}:
\byear{1953},
\batitle{{The Solar Magnetograph.}}
\bjtitle{\apj}
\bvolume{118},
\bfpage{387}.
\doiurl{https://doi.org/10.1086/145767}.
\adsurl{1953ApJ...118..387B}.
\end{barticle}
\endbibitem

\bibitem[\protect\citeauthoryear{{Beck} et~al.}{2005a}]{Beck2005}
\begin{barticle}
\bauthor{\bsnm{{Beck}}, \binits{C.}},
\bauthor{\bsnm{{Schlichenmaier}}, \binits{R.}},
\bauthor{\bsnm{{Collados}}, \binits{M.}},
\bauthor{\bsnm{{Bellot Rubio}}, \binits{L.}},
\bauthor{\bsnm{{Kentischer}}, \binits{T.}}:
\byear{2005}a,
\batitle{{A polarization model for the German Vacuum Tower Telescope from in
  situ and laboratory measurements}}.
\bjtitle{\aap}
\bvolume{443},
\bfpage{1047}.
\doiurl{https://doi.org/10.1051/0004-6361:20052935}.
\adsurl{2005A&A...443.1047B}.
\end{barticle}
\endbibitem

\bibitem[\protect\citeauthoryear{{Beck} et~al.}{2005b}]{Beck2005b}
\begin{barticle}
\bauthor{\bsnm{{Beck}}, \binits{C.}},
\bauthor{\bsnm{{Schlichenmaier}}, \binits{R.}},
\bauthor{\bsnm{{Collados}}, \binits{M.}},
\bauthor{\bsnm{{Bellot Rubio}}, \binits{L.}},
\bauthor{\bsnm{{Kentischer}}, \binits{T.}}:
\byear{2005}b,
\batitle{{A polarization model for the German Vacuum Tower Telescope from in
  situ and laboratory measurements}}.
\bjtitle{\aap}
\bvolume{443},
\bfpage{1047}.
\doiurl{https://doi.org/10.1051/0004-6361:20052935}.
\adsurl{2005A&A...443.1047B}.
\end{barticle}
\endbibitem

\bibitem[\protect\citeauthoryear{{Cao} et~al.}{2010}]{Cao2010}
\begin{barticle}
\bauthor{\bsnm{{Cao}}, \binits{W.}},
\bauthor{\bsnm{{Gorceix}}, \binits{N.}},
\bauthor{\bsnm{{Coulter}}, \binits{R.}},
\bauthor{\bsnm{{Ahn}}, \binits{K.}},
\bauthor{\bsnm{{Rimmele}}, \binits{T.R.}},
\bauthor{\bsnm{{Goode}}, \binits{P.R.}}:
\byear{2010},
\batitle{{Scientific instrumentation for the 1.6 m New Solar Telescope in Big
  Bear}}.
\bjtitle{Astron. Nachr.}
\bvolume{331},
\bfpage{636}.
\doiurl{https://doi.org/10.1002/asna.201011390}.
\adsurl{2010AN....331..636C}.
\end{barticle}
\endbibitem

\bibitem[\protect\citeauthoryear{{Cauzzi} et~al.}{2008}]{Cauzzi2008}
\begin{barticle}
\bauthor{\bsnm{{Cauzzi}}, \binits{G.}},
\bauthor{\bsnm{{Reardon}}, \binits{K.P.}},
\bauthor{\bsnm{{Uitenbroek}}, \binits{H.}},
\bauthor{\bsnm{{Cavallini}}, \binits{F.}},
\bauthor{\bsnm{{Falchi}}, \binits{A.}},
\bauthor{\bsnm{{Falciani}}, \binits{R.}},
\bauthor{\bsnm{{Janssen}}, \binits{K.}},
\bauthor{\bsnm{{Rimmele}}, \binits{T.}},
\bauthor{\bsnm{{Vecchio}}, \binits{A.}},
\bauthor{\bsnm{{W{\"o}ger}}, \binits{F.}}:
\byear{2008},
\batitle{{The solar chromosphere at high resolution with IBIS. I. New insights
  from the Ca II 854.2 nm line}}.
\bjtitle{\aap}
\bvolume{480},
\bfpage{515}.
\doiurl{https://doi.org/10.1051/0004-6361:20078642}.
\adsurl{2008A&A...480..515C}.
\end{barticle}
\endbibitem

\bibitem[\protect\citeauthoryear{{Collados} et~al.}{2012}]{Collados2012}
\begin{barticle}
\bauthor{\bsnm{{Collados}}, \binits{M.}},
\bauthor{\bsnm{{L{\'o}pez}}, \binits{R.}},
\bauthor{\bsnm{{P{\'a}ez}}, \binits{E.}},
\bauthor{\bsnm{{Hern{\'a}ndez}}, \binits{E.}},
\bauthor{\bsnm{{Reyes}}, \binits{M.}},
\bauthor{\bsnm{{Calcines}}, \binits{A.}},
\bauthor{\bsnm{{Ballesteros}}, \binits{E.}},
\bauthor{\bsnm{{D{\'\i}az}}, \binits{J.J.}},
\bauthor{\bsnm{{Denker}}, \binits{C.}},
\bauthor{\bsnm{{Lagg}}, \binits{A.}},
\bauthor{\bsnm{{Schlichenmaier}}, \binits{R.}},
\bauthor{\bsnm{{Schmidt}}, \binits{W.}},
\bauthor{\bsnm{{Solanki}}, \binits{S.K.}},
\bauthor{\bsnm{{Strassmeier}}, \binits{K.G.}},
\bauthor{\bsnm{{von der L{\"u}he}}, \binits{O.}},
\bauthor{\bsnm{{Volkmer}}, \binits{R.}}:
\byear{2012},
\batitle{{GRIS: The GREGOR Infrared Spectrograph}}.
\bjtitle{Astron. Nachr.}
\bvolume{333},
\bfpage{872}.
\doiurl{https://doi.org/10.1002/asna.201211738}.
\adsurl{2012AN....333..872C}.
\end{barticle}
\endbibitem

\bibitem[\protect\citeauthoryear{{Collados} et~al.}{2013}]{Collados2013}
\begin{barticle}
\bauthor{\bsnm{{Collados}}, \binits{M.}},
\bauthor{\bsnm{{Bettonvil}}, \binits{F.}},
\bauthor{\bsnm{{Cavaller}}, \binits{L.}},
\bauthor{\bsnm{{Ermolli}}, \binits{I.}},
\bauthor{\bsnm{{Gelly}}, \binits{B.}},
\bauthor{\bsnm{{P{\'e}rez}}, \binits{A.}},
\bauthor{\bsnm{{Socas-Navarro}}, \binits{H.}},
\bauthor{\bsnm{{Soltau}}, \binits{D.}},
\bauthor{\bsnm{{Volkmer}}, \binits{R.}},
\bauthor{\bsnm{{EST Team}}}:
\byear{2013},
\batitle{{The European Solar Telescope}}.
\bjtitle{Memorie della Societa Astronomica Italiana}
\bvolume{84},
\bfpage{379}.
\adsurl{2013MmSAI..84..379C}.
\end{barticle}
\endbibitem

\bibitem[\protect\citeauthoryear{{de Wijn} et~al.}{2022}]{deWijn2022}
\begin{barticle}
\bauthor{\bsnm{{de Wijn}}, \binits{A.G.}},
\bauthor{\bsnm{{Casini}}, \binits{R.}},
\bauthor{\bsnm{{Carlile}}, \binits{A.}},
\bauthor{\bsnm{{Lecinski}}, \binits{A.R.}},
\bauthor{\bsnm{{Sewell}}, \binits{S.}},
\bauthor{\bsnm{{Zmarzly}}, \binits{P.}},
\bauthor{\bsnm{{Eigenbrot}}, \binits{A.D.}},
\bauthor{\bsnm{{Beck}}, \binits{C.}},
\bauthor{\bsnm{{W{\"o}ger}}, \binits{F.}},
\bauthor{\bsnm{{Kn{\"o}lker}}, \binits{M.}}:
\byear{2022},
\batitle{{The Visible Spectro-Polarimeter of the Daniel K. Inouye Solar
  Telescope}}.
\bjtitle{\solphys}
\bvolume{297},
\bfpage{22}.
\doiurl{https://doi.org/10.1007/s11207-022-01954-1}.
\adsurl{2022SoPh..297...22D}.
\end{barticle}
\endbibitem

\bibitem[\protect\citeauthoryear{{del Toro Iniesta}}{2003}]{delToroIniesta2003}
\begin{bbook}
\bauthor{\bsnm{{del Toro Iniesta}}, \binits{J.C.}}:
\byear{2003},
\bbtitle{{Introduction to Spectropolarimetry}}.
\adsurl{2003isp..book.....D}.
\end{bbook}
\endbibitem

\bibitem[\protect\citeauthoryear{{D{\'\i}az Baso} et~al.}{2019}]{DiazBaso2019}
\begin{barticle}
\bauthor{\bsnm{{D{\'\i}az Baso}}, \binits{C.J.}},
\bauthor{\bsnm{{Mart{\'\i}nez Gonz{\'a}lez}}, \binits{M.J.}},
\bauthor{\bsnm{{Asensio Ramos}}, \binits{A.}},
\bauthor{\bsnm{{de la Cruz Rodr{\'\i}guez}}, \binits{J.}}:
\byear{2019},
\batitle{{Diagnostic potential of the Ca II 8542 {\r{A}} line for solar
  filaments}}.
\bjtitle{\aap}
\bvolume{623},
\bfpage{A178}.
\doiurl{https://doi.org/10.1051/0004-6361/201834793}.
\adsurl{2019A&A...623A.178D}.
\end{barticle}
\endbibitem

\bibitem[\protect\citeauthoryear{{Eversberg} and
  {Vollmann}}{2015}]{Eversberg2015}
\begin{bbook}
\bauthor{\bsnm{{Eversberg}}, \binits{T.}},
\bauthor{\bsnm{{Vollmann}}, \binits{K.}}:
\byear{2015},
\bbtitle{{Spectroscopic Instrumentation: Fundamentals and Guidelines for
  Astronomers}}.
\doiurl{https://doi.org/10.1007/978-3-662-44535-8}.
\adsurl{2015sifg.book.....E}.
\end{bbook}
\endbibitem

\bibitem[\protect\citeauthoryear{{Fehlmann} et~al.}{2023}]{Fehlmann2023}
\begin{barticle}
\bauthor{\bsnm{{Fehlmann}}, \binits{A.}},
\bauthor{\bsnm{{Kuhn}}, \binits{J.R.}},
\bauthor{\bsnm{{Schad}}, \binits{T.A.}},
\bauthor{\bsnm{{Scholl}}, \binits{I.F.}},
\bauthor{\bsnm{{Williams}}, \binits{R.}},
\bauthor{\bsnm{{Agdinaoay}}, \binits{R.}},
\bauthor{\bsnm{{Berst}}, \binits{D.C.}},
\bauthor{\bsnm{{Craig}}, \binits{S.C.}},
\bauthor{\bsnm{{Giebink}}, \binits{C.}},
\bauthor{\bsnm{{Goodrich}}, \binits{B.}},
\bauthor{\bsnm{{Hnat}}, \binits{K.}},
\bauthor{\bsnm{{James}}, \binits{D.}},
\bauthor{\bsnm{{Lockhart}}, \binits{C.}},
\bauthor{\bsnm{{Mickey}}, \binits{D.L.}},
\bauthor{\bsnm{{Oswald}}, \binits{D.}},
\bauthor{\bsnm{{Puentes}}, \binits{M.M.}},
\bauthor{\bsnm{{Schickling}}, \binits{R.}},
\bauthor{\bsnm{{de Vanssay}}, \binits{J.-B.}},
\bauthor{\bsnm{{Warmbier}}, \binits{E.A.}}:
\byear{2023},
\batitle{{The Daniel K. Inouye Solar Telescope (DKIST) Cryogenic Near-Infrared
  Spectro-Polarimeter}}.
\bjtitle{\solphys}
\bvolume{298},
\bfpage{5}.
\doiurl{https://doi.org/10.1007/s11207-022-02098-y}.
\adsurl{2023SoPh..298....5F}.
\end{barticle}
\endbibitem

\bibitem[\protect\citeauthoryear{{Galazutdinov}}{2022}]{Galazutdinov2022}
\begin{barticle}
\bauthor{\bsnm{{Galazutdinov}}, \binits{G.A.}}:
\byear{2022},
\batitle{{DECH: A Software Package for Astronomical Spectral Data Processing
  and Analysis}}.
\bjtitle{Astrophys. Bull.}
\bvolume{77},
\bfpage{519}.
\doiurl{https://doi.org/10.1134/S1990341322040034}.
\adsurl{2022AstBu..77..519G}.
\end{barticle}
\endbibitem

\bibitem[\protect\citeauthoryear{{Haneychuk}, {Kotov}, and
  {Tsap}}{2003}]{Haneychuk2003}
\begin{barticle}
\bauthor{\bsnm{{Haneychuk}}, \binits{V.I.}},
\bauthor{\bsnm{{Kotov}}, \binits{V.A.}},
\bauthor{\bsnm{{Tsap}}, \binits{T.T.}}:
\byear{2003},
\batitle{{On stability of rotation of the mean magnetic field of the Sun}}.
\bjtitle{\aap}
\bvolume{403},
\bfpage{1115}.
\doiurl{https://doi.org/10.1051/0004-6361:20030426}.
\adsurl{2003A&A...403.1115H}.
\end{barticle}
\endbibitem

\bibitem[\protect\citeauthoryear{{Ichimoto} et~al.}{2008}]{Ichimoto2008}
\begin{barticle}
\bauthor{\bsnm{{Ichimoto}}, \binits{K.}},
\bauthor{\bsnm{{Lites}}, \binits{B.}},
\bauthor{\bsnm{{Elmore}}, \binits{D.}},
\bauthor{\bsnm{{Suematsu}}, \binits{Y.}},
\bauthor{\bsnm{{Tsuneta}}, \binits{S.}},
\bauthor{\bsnm{{Katsukawa}}, \binits{Y.}},
\bauthor{\bsnm{{Shimizu}}, \binits{T.}},
\bauthor{\bsnm{{Shine}}, \binits{R.}},
\bauthor{\bsnm{{Tarbell}}, \binits{T.}},
\bauthor{\bsnm{{Title}}, \binits{A.}},
\bauthor{\bsnm{{Kiyohara}}, \binits{J.}},
\bauthor{\bsnm{{Shinoda}}, \binits{K.}},
\bauthor{\bsnm{{Card}}, \binits{G.}},
\bauthor{\bsnm{{Lecinski}}, \binits{A.}},
\bauthor{\bsnm{{Streander}}, \binits{K.}},
\bauthor{\bsnm{{Nakagiri}}, \binits{M.}},
\bauthor{\bsnm{{Miyashita}}, \binits{M.}},
\bauthor{\bsnm{{Noguchi}}, \binits{M.}},
\bauthor{\bsnm{{Hoffmann}}, \binits{C.}},
\bauthor{\bsnm{{Cruz}}, \binits{T.}}:
\byear{2008},
\batitle{{Polarization Calibration of the Solar Optical Telescope onboard
  Hinode}}.
\bjtitle{\solphys}
\bvolume{249},
\bfpage{233}.
\doiurl{https://doi.org/10.1007/s11207-008-9169-9}.
\adsurl{2008SoPh..249..233I}.
\end{barticle}
\endbibitem

\bibitem[\protect\citeauthoryear{{Jaeggli} et~al.}{2022}]{Jaeggli2022}
\begin{barticle}
\bauthor{\bsnm{{Jaeggli}}, \binits{S.A.}},
\bauthor{\bsnm{{Lin}}, \binits{H.}},
\bauthor{\bsnm{{Onaka}}, \binits{P.}},
\bauthor{\bsnm{{Yamada}}, \binits{H.}},
\bauthor{\bsnm{{Anan}}, \binits{T.}},
\bauthor{\bsnm{{Bonnet}}, \binits{M.}},
\bauthor{\bsnm{{Ching}}, \binits{G.}},
\bauthor{\bsnm{{Huang}}, \binits{X.-P.}},
\bauthor{\bsnm{{Kramar}}, \binits{M.}},
\bauthor{\bsnm{{McGregor}}, \binits{H.}},
\bauthor{\bsnm{{Nitta}}, \binits{G.}},
\bauthor{\bsnm{{Rae}}, \binits{C.}},
\bauthor{\bsnm{{Robertson}}, \binits{L.}},
\bauthor{\bsnm{{Schad}}, \binits{T.A.}},
\bauthor{\bsnm{{Toyama}}, \binits{P.}},
\bauthor{\bsnm{{Young}}, \binits{J.}},
\bauthor{\bsnm{{Berst}}, \binits{C.}},
\bauthor{\bsnm{{Harrington}}, \binits{D.M.}},
\bauthor{\bsnm{{Liang}}, \binits{M.}},
\bauthor{\bsnm{{Puentes}}, \binits{M.}},
\bauthor{\bsnm{{Sekulic}}, \binits{P.}},
\bauthor{\bsnm{{Smith}}, \binits{B.}},
\bauthor{\bsnm{{Sueoka}}, \binits{S.R.}}:
\byear{2022},
\batitle{{The Diffraction-Limited Near-Infrared Spectropolarimeter (DL-NIRSP)
  of the Daniel K. Inouye Solar Telescope (DKIST)}}.
\bjtitle{\solphys}
\bvolume{297},
\bfpage{137}.
\doiurl{https://doi.org/10.1007/s11207-022-02062-w}.
\adsurl{2022SoPh..297..137J}.
\end{barticle}
\endbibitem

\bibitem[\protect\citeauthoryear{{Jarolim} et~al.}{2024}]{Jarolim2024}
\begin{barticle}
\bauthor{\bsnm{{Jarolim}}, \binits{R.}},
\bauthor{\bsnm{{Tremblay}}, \binits{B.}},
\bauthor{\bsnm{{Rempel}}, \binits{M.}},
\bauthor{\bsnm{{Molnar}}, \binits{M.}},
\bauthor{\bsnm{{Veronig}}, \binits{A.M.}},
\bauthor{\bsnm{{Thalmann}}, \binits{J.K.}},
\bauthor{\bsnm{{Podladchikova}}, \binits{T.}}:
\byear{2024},
\batitle{{Advancing Solar Magnetic Field Extrapolations through Multiheight
  Magnetic Field Measurements}}.
\bjtitle{\apjl}
\bvolume{963},
\bfpage{L21}.
\doiurl{https://doi.org/10.3847/2041-8213/ad2450}.
\adsurl{2024ApJ...963L..21J}.
\end{barticle}
\endbibitem

\bibitem[\protect\citeauthoryear{{Judge} et~al.}{2004}]{Judge2004}
\begin{barticle}
\bauthor{\bsnm{{Judge}}, \binits{P.G.}},
\bauthor{\bsnm{{Elmore}}, \binits{D.F.}},
\bauthor{\bsnm{{Lites}}, \binits{B.W.}},
\bauthor{\bsnm{{Keller}}, \binits{C.U.}},
\bauthor{\bsnm{{Rimmele}}, \binits{T.}}:
\byear{2004},
\batitle{{Evaluation of Seeing-Induced Cross Talk in Tip-Tilt-Corrected Solar
  Polarimetry}}.
\bjtitle{\ao}
\bvolume{43},
\bfpage{3817}.
\doiurl{https://doi.org/10.1364/AO.43.003817}.
\adsurl{2004ApOpt..43.3817J}.
\end{barticle}
\endbibitem

\bibitem[\protect\citeauthoryear{{Keller}, {Harvey}, and
  {Giampapa}}{2003}]{Keller2003}
\begin{bchapter}
\bauthor{\bsnm{{Keller}}, \binits{C.U.}},
\bauthor{\bsnm{{Harvey}}, \binits{J.W.}},
\bauthor{\bsnm{{Giampapa}}, \binits{M.S.}}:
\byear{2003},
\bctitle{{SOLIS: an innovative suite of synoptic instruments}}.
In: \beditor{\bsnm{{Keil}}, \binits{S.L.}},
\beditor{\bsnm{{Avakyan}}, \binits{S.V.}} (eds.)
\bbtitle{Innovative Telescopes and Instrumentation for Solar Astrophysics},
\bsertitle{Society of Photo-Optical Instrumentation Engineers (SPIE) Conference
  Series}
\bseriesno{4853},
\bfpage{194}.
\doiurl{https://doi.org/10.1117/12.460373}.
\adsurl{2003SPIE.4853..194K}.
\end{bchapter}
\endbibitem

\bibitem[\protect\citeauthoryear{{Kotov}, {Severny}, and
  {Tsap}}{1982}]{Kotov1982}
\begin{barticle}
\bauthor{\bsnm{{Kotov}}, \binits{V.A.}},
\bauthor{\bsnm{{Severny}}, \binits{A.B.}},
\bauthor{\bsnm{{Tsap}}, \binits{T.T.}}:
\byear{1982},
\batitle{{Study of global oscillations of the Sun: I. The method and the
  instrument}}.
\bjtitle{Izv. Krymsk. Astrofiz. Observ. (in Russian)}
\bvolume{65},
\bfpage{3}.
\end{barticle}
\endbibitem

\bibitem[\protect\citeauthoryear{{Kutsenko} et~al.}{2022}]{Kutsenko2022}
\begin{barticle}
\bauthor{\bsnm{{Kutsenko}}, \binits{A.S.}},
\bauthor{\bsnm{{Yang}}, \binits{S.}},
\bauthor{\bsnm{{Abramenko}}, \binits{V.}},
\bauthor{\bsnm{{Semyonov}}, \binits{D.}}:
\byear{2022},
\batitle{{Experiments on high-spatial-resolution observations of the Sun at the
  A.B. Severny Solar Tower Telescope of the Crimean Astrophysical
  Observatory}}.
\bjtitle{Acta Astrophys. Taurica}
\bvolume{3},
\bfpage{8}.
\doiurl{https://doi.org/10.34898/aat.vol3.iss3.pp8-11}.
\adsurl{2022AcAT....3c...8K}.
\end{barticle}
\endbibitem

\bibitem[\protect\citeauthoryear{{Lagg} et~al.}{2017}]{Lagg2017}
\begin{barticle}
\bauthor{\bsnm{{Lagg}}, \binits{A.}},
\bauthor{\bsnm{{Lites}}, \binits{B.}},
\bauthor{\bsnm{{Harvey}}, \binits{J.}},
\bauthor{\bsnm{{Gosain}}, \binits{S.}},
\bauthor{\bsnm{{Centeno}}, \binits{R.}}:
\byear{2017},
\batitle{{Measurements of Photospheric and Chromospheric Magnetic Fields}}.
\bjtitle{\ssr}
\bvolume{210},
\bfpage{37}.
\doiurl{https://doi.org/10.1007/s11214-015-0219-y}.
\adsurl{2017SSRv..210...37L}.
\end{barticle}
\endbibitem

\bibitem[\protect\citeauthoryear{{Landi Degl'Innocenti} and
  {Landolfi}}{2004}]{Landi2004}
\begin{bbook}
\bauthor{\bsnm{{Landi Degl'Innocenti}}, \binits{E.}},
\bauthor{\bsnm{{Landolfi}}, \binits{M.}}:
\byear{2004},
\bbtitle{{Polarization in Spectral Lines}}
\bseriesno{307}.
\doiurl{https://doi.org/10.1007/978-1-4020-2415-3}.
\adsurl{2004ASSL..307.....L}.
\end{bbook}
\endbibitem

\bibitem[\protect\citeauthoryear{{Lites}}{1987}]{Lites1987}
\begin{barticle}
\bauthor{\bsnm{{Lites}}, \binits{B.W.}}:
\byear{1987},
\batitle{{Rotating waveplates as polarization modulators for Stokes polarimetry
  of the sun: evaluation of seeing-induced crosstalk errors}}.
\bjtitle{\ao}
\bvolume{26},
\bfpage{3838}.
\doiurl{https://doi.org/10.1364/AO.26.003838}.
\adsurl{1987ApOpt..26.3838L}.
\end{barticle}
\endbibitem

\bibitem[\protect\citeauthoryear{{Lites} and {Ichimoto}}{2013}]{Lites2013}
\begin{barticle}
\bauthor{\bsnm{{Lites}}, \binits{B.W.}},
\bauthor{\bsnm{{Ichimoto}}, \binits{K.}}:
\byear{2013},
\batitle{{The SP\_PREP Data Preparation Package for the Hinode
  Spectro-Polarimeter}}.
\bjtitle{\solphys}
\bvolume{283},
\bfpage{601}.
\doiurl{https://doi.org/10.1007/s11207-012-0205-4}.
\adsurl{2013SoPh..283..601L}.
\end{barticle}
\endbibitem

\bibitem[\protect\citeauthoryear{{Pasachoff} and
  {Livingston}}{1984}]{Pasachoff1984}
\begin{barticle}
\bauthor{\bsnm{{Pasachoff}}, \binits{J.M.}},
\bauthor{\bsnm{{Livingston}}, \binits{W.C.}}:
\byear{1984},
\batitle{{Coelostat and heliostat: alignment and use for eclipse and other
  field purposes}}.
\bjtitle{Applied Optics}
\bvolume{23},
\bfpage{2803}.
\doiurl{https://doi.org/10.1364/AO.23.002803}.
\adsurl{1984ApOpt..23.2803P}.
\end{barticle}
\endbibitem

\bibitem[\protect\citeauthoryear{{Pastor Yabar} et~al.}{2021}]{PastorYabar2021}
\begin{barticle}
\bauthor{\bsnm{{Pastor Yabar}}, \binits{A.}},
\bauthor{\bsnm{{Borrero}}, \binits{J.M.}},
\bauthor{\bsnm{{Quintero Noda}}, \binits{C.}},
\bauthor{\bsnm{{Ruiz Cobo}}, \binits{B.}}:
\byear{2021},
\batitle{{Inference of electric currents in the solar photosphere}}.
\bjtitle{\aap}
\bvolume{656},
\bfpage{L20}.
\doiurl{https://doi.org/10.1051/0004-6361/202142149}.
\adsurl{2021A&A...656L..20P}.
\end{barticle}
\endbibitem

\bibitem[\protect\citeauthoryear{{Pesnell}, {Thompson}, and
  {Chamberlin}}{2012}]{Pesnell2012}
\begin{barticle}
\bauthor{\bsnm{{Pesnell}}, \binits{W.D.}},
\bauthor{\bsnm{{Thompson}}, \binits{B.J.}},
\bauthor{\bsnm{{Chamberlin}}, \binits{P.C.}}:
\byear{2012},
\batitle{{The Solar Dynamics Observatory (SDO)}}.
\bjtitle{\solphys}
\bvolume{275},
\bfpage{3}.
\doiurl{https://doi.org/10.1007/s11207-011-9841-3}.
\adsurl{2012SoPh..275....3P}.
\end{barticle}
\endbibitem

\bibitem[\protect\citeauthoryear{{Piskunov} and {Valenti}}{2002}]{Piskunov2002}
\begin{barticle}
\bauthor{\bsnm{{Piskunov}}, \binits{N.E.}},
\bauthor{\bsnm{{Valenti}}, \binits{J.A.}}:
\byear{2002},
\batitle{{New algorithms for reducing cross-dispersed echelle spectra}}.
\bjtitle{\aap}
\bvolume{385},
\bfpage{1095}.
\doiurl{https://doi.org/10.1051/0004-6361:20020175}.
\adsurl{2002A&A...385.1095P}.
\end{barticle}
\endbibitem

\bibitem[\protect\citeauthoryear{{Piskunov}, {Wehrhahn}, and
  {Marquart}}{2021}]{Piskunov2021}
\begin{barticle}
\bauthor{\bsnm{{Piskunov}}, \binits{N.}},
\bauthor{\bsnm{{Wehrhahn}}, \binits{A.}},
\bauthor{\bsnm{{Marquart}}, \binits{T.}}:
\byear{2021},
\batitle{{Optimal extraction of echelle spectra: Getting the most out of
  observations}}.
\bjtitle{\aap}
\bvolume{646},
\bfpage{A32}.
\doiurl{https://doi.org/10.1051/0004-6361/202038293}.
\adsurl{2021A&A...646A..32P}.
\end{barticle}
\endbibitem

\bibitem[\protect\citeauthoryear{{Pruthvi} and {Roth}}{2023}]{Pruthvi2023}
\begin{barticle}
\bauthor{\bsnm{{Pruthvi}}, \binits{H.}},
\bauthor{\bsnm{{Roth}}, \binits{M.}}:
\byear{2023},
\batitle{{The New HELLRIDE at the Vacuum Tower Telescope}}.
\bjtitle{\solphys}
\bvolume{298},
\bfpage{41}.
\doiurl{https://doi.org/10.1007/s11207-023-02138-1}.
\adsurl{2023SoPh..298...41P}.
\end{barticle}
\endbibitem

\bibitem[\protect\citeauthoryear{{Quintero Noda}
  et~al.}{2018}]{QuinteroNoda2018}
\begin{barticle}
\bauthor{\bsnm{{Quintero Noda}}, \binits{C.}},
\bauthor{\bsnm{{Uitenbroek}}, \binits{H.}},
\bauthor{\bsnm{{Carlsson}}, \binits{M.}},
\bauthor{\bsnm{{Orozco Su{\'a}rez}}, \binits{D.}},
\bauthor{\bsnm{{Katsukawa}}, \binits{Y.}},
\bauthor{\bsnm{{Shimizu}}, \binits{T.}},
\bauthor{\bsnm{{Ruiz Cobo}}, \binits{B.}},
\bauthor{\bsnm{{Kubo}}, \binits{M.}},
\bauthor{\bsnm{{Oba}}, \binits{T.}},
\bauthor{\bsnm{{Kawabata}}, \binits{Y.}},
\bauthor{\bsnm{{Hasegawa}}, \binits{T.}},
\bauthor{\bsnm{{Ichimoto}}, \binits{K.}},
\bauthor{\bsnm{{Anan}}, \binits{T.}},
\bauthor{\bsnm{{Suematsu}}, \binits{Y.}}:
\byear{2018},
\batitle{{Study of the polarization produced by the Zeeman effect in the solar
  Mg I b lines}}.
\bjtitle{\mnras}
\bvolume{481},
\bfpage{5675}.
\doiurl{https://doi.org/10.1093/mnras/sty2685}.
\adsurl{2018MNRAS.481.5675Q}.
\end{barticle}
\endbibitem

\bibitem[\protect\citeauthoryear{{Quintero Noda}
  et~al.}{2022}]{QuinteroNoda2022}
\begin{barticle}
\bauthor{\bsnm{{Quintero Noda}}, \binits{C.}},
\bauthor{\bsnm{{Schlichenmaier}}, \binits{R.}},
\bauthor{\bsnm{{Bellot Rubio}}, \binits{L.R.}},
\bauthor{\bsnm{{L{\"o}fdahl}}, \binits{M.G.}},
\bauthor{\bsnm{{Khomenko}}, \binits{E.}},
\bauthor{\bsnm{{Jur{\v{c}}{\'a}k}}, \binits{J.}},
\bauthor{\bsnm{{Leenaarts}}, \binits{J.}},
\bauthor{\bsnm{{Kuckein}}, \binits{C.}},
\bauthor{\bsnm{{Gonz{\'a}lez Manrique}}, \binits{S.J.}},
\bauthor{\bsnm{{Gun{\'a}r}}, \binits{S.}},
\bauthor{\bsnm{{Nelson}}, \binits{C.J.}},
\bauthor{\bsnm{{de la Cruz Rodr{\'\i}guez}}, \binits{J.}},
\bauthor{\bsnm{{Tziotziou}}, \binits{K.}},
\bauthor{\bsnm{{Tsiropoula}}, \binits{G.}},
\bauthor{\bsnm{{Aulanier}}, \binits{G.}},
\bauthor{\bsnm{{Aboudarham}}, \binits{J.}},
\bauthor{\bsnm{{Allegri}}, \binits{D.}},
\bauthor{\bsnm{{Alsina Ballester}}, \binits{E.}},
\bauthor{\bsnm{{Amans}}, \binits{J.P.}},
\bauthor{\bsnm{{Asensio Ramos}}, \binits{A.}},
\bauthor{\bsnm{{Bail{\'e}n}}, \binits{F.J.}},
\bauthor{\bsnm{{Balaguer}}, \binits{M.}},
\bauthor{\bsnm{{Baldini}}, \binits{V.}},
\bauthor{\bsnm{{Balthasar}}, \binits{H.}},
\bauthor{\bsnm{{Barata}}, \binits{T.}},
\bauthor{\bsnm{{Barczynski}}, \binits{K.}},
\bauthor{\bsnm{{Barreto Cabrera}}, \binits{M.}},
\bauthor{\bsnm{{Baur}}, \binits{A.}},
\bauthor{\bsnm{{B{\'e}chet}}, \binits{C.}},
\bauthor{\bsnm{{Beck}}, \binits{C.}},
\bauthor{\bsnm{{Bel{\'\i}o-As{\'\i}n}}, \binits{M.}},
\bauthor{\bsnm{{Bello-Gonz{\'a}lez}}, \binits{N.}},
\bauthor{\bsnm{{Belluzzi}}, \binits{L.}},
\bauthor{\bsnm{{Bentley}}, \binits{R.D.}},
\bauthor{\bsnm{{Berdyugina}}, \binits{S.V.}},
\bauthor{\bsnm{{Berghmans}}, \binits{D.}},
\bauthor{\bsnm{{Berlicki}}, \binits{A.}},
\bauthor{\bsnm{{Berrilli}}, \binits{F.}},
\bauthor{\bsnm{{Berkefeld}}, \binits{T.}},
\bauthor{\bsnm{{Bettonvil}}, \binits{F.}},
\bauthor{\bsnm{{Bianda}}, \binits{M.}},
\bauthor{\bsnm{{Bienes P{\'e}rez}}, \binits{J.}},
\bauthor{\bsnm{{Bonaque-Gonz{\'a}lez}}, \binits{S.}},
\bauthor{\bsnm{{Braj{\v{s}}a}}, \binits{R.}},
\bauthor{\bsnm{{Bommier}}, \binits{V.}},
\bauthor{\bsnm{{Bourdin}}, \binits{P.-A.}},
\bauthor{\bsnm{{Burgos Mart{\'\i}n}}, \binits{J.}},
\bauthor{\bsnm{{Calchetti}}, \binits{D.}},
\bauthor{\bsnm{{Calcines}}, \binits{A.}},
\bauthor{\bsnm{{Calvo Tovar}}, \binits{J.}},
\bauthor{\bsnm{{Campbell}}, \binits{R.J.}},
\bauthor{\bsnm{{Carballo-Mart{\'\i}n}}, \binits{Y.}},
\bauthor{\bsnm{{Carbone}}, \binits{V.}},
\bauthor{\bsnm{{Carlin}}, \binits{E.S.}},
\bauthor{\bsnm{{Carlsson}}, \binits{M.}},
\bauthor{\bsnm{{Castro L{\'o}pez}}, \binits{J.}},
\bauthor{\bsnm{{Cavaller}}, \binits{L.}},
\bauthor{\bsnm{{Cavallini}}, \binits{F.}},
\bauthor{\bsnm{{Cauzzi}}, \binits{G.}},
\bauthor{\bsnm{{Cecconi}}, \binits{M.}},
\bauthor{\bsnm{{Chulani}}, \binits{H.M.}},
\bauthor{\bsnm{{Cirami}}, \binits{R.}},
\bauthor{\bsnm{{Consolini}}, \binits{G.}},
\bauthor{\bsnm{{Coretti}}, \binits{I.}},
\bauthor{\bsnm{{Cosentino}}, \binits{R.}},
\bauthor{\bsnm{{C{\'o}zar-Castellano}}, \binits{J.}},
\bauthor{\bsnm{{Dalmasse}}, \binits{K.}},
\bauthor{\bsnm{{Danilovic}}, \binits{S.}},
\bauthor{\bsnm{{De Juan Ovelar}}, \binits{M.}},
\bauthor{\bsnm{{Del Moro}}, \binits{D.}},
\bauthor{\bsnm{{del Pino Alem{\'a}n}}, \binits{T.}},
\bauthor{\bsnm{{del Toro Iniesta}}, \binits{J.C.}},
\bauthor{\bsnm{{Denker}}, \binits{C.}},
\bauthor{\bsnm{{Dhara}}, \binits{S.K.}},
\bauthor{\bsnm{{Di Marcantonio}}, \binits{P.}},
\bauthor{\bsnm{{D{\'\i}az Baso}}, \binits{C.J.}},
\bauthor{\bsnm{{Diercke}}, \binits{A.}},
\bauthor{\bsnm{{Dineva}}, \binits{E.}},
\bauthor{\bsnm{{D{\'\i}az-Garc{\'\i}a}}, \binits{J.J.}},
\bauthor{\bsnm{{Doerr}}, \binits{H.-P.}},
\bauthor{\bsnm{{Doyle}}, \binits{G.}},
\bauthor{\bsnm{{Erdelyi}}, \binits{R.}},
\bauthor{\bsnm{{Ermolli}}, \binits{I.}},
\bauthor{\bsnm{{Escobar Rodr{\'\i}guez}}, \binits{A.}},
\bauthor{\bsnm{{Esteban Pozuelo}}, \binits{S.}},
\bauthor{\bsnm{{Faurobert}}, \binits{M.}},
\bauthor{\bsnm{{Felipe}}, \binits{T.}},
\bauthor{\bsnm{{Feller}}, \binits{A.}},
\bauthor{\bsnm{{Feijoo Amoedo}}, \binits{N.}},
\bauthor{\bsnm{{Femen{\'\i}a Castell{\'a}}}, \binits{B.}},
\bauthor{\bsnm{{Fernandes}}, \binits{J.}},
\bauthor{\bsnm{{Ferro Rodr{\'\i}guez}}, \binits{I.}},
\bauthor{\bsnm{{Figueroa}}, \binits{I.}},
\bauthor{\bsnm{{Fletcher}}, \binits{L.}},
\bauthor{\bsnm{{Franco Ordovas}}, \binits{A.}},
\bauthor{\bsnm{{Gafeira}}, \binits{R.}},
\bauthor{\bsnm{{Gardenghi}}, \binits{R.}},
\bauthor{\bsnm{{Gelly}}, \binits{B.}},
\bauthor{\bsnm{{Giorgi}}, \binits{F.}},
\bauthor{\bsnm{{Gisler}}, \binits{D.}},
\bauthor{\bsnm{{Giovannelli}}, \binits{L.}},
\bauthor{\bsnm{{Gonz{\'a}lez}}, \binits{F.}},
\bauthor{\bsnm{{Gonz{\'a}lez}}, \binits{J.B.}},
\bauthor{\bsnm{{Gonz{\'a}lez-Cava}}, \binits{J.M.}},
\bauthor{\bsnm{{Gonz{\'a}lez Garc{\'\i}a}}, \binits{M.}},
\bauthor{\bsnm{{G{\"o}m{\"o}ry}}, \binits{P.}},
\bauthor{\bsnm{{Gracia}}, \binits{F.}},
\bauthor{\bsnm{{Grauf}}, \binits{B.}},
\bauthor{\bsnm{{Greco}}, \binits{V.}},
\bauthor{\bsnm{{Grivel}}, \binits{C.}},
\bauthor{\bsnm{{Guerreiro}}, \binits{N.}},
\bauthor{\bsnm{{Guglielmino}}, \binits{S.L.}},
\bauthor{\bsnm{{Hammerschlag}}, \binits{R.}},
\bauthor{\bsnm{{Hanslmeier}}, \binits{A.}},
\bauthor{\bsnm{{Hansteen}}, \binits{V.}},
\bauthor{\bsnm{{Heinzel}}, \binits{P.}},
\bauthor{\bsnm{{Hern{\'a}ndez-Delgado}}, \binits{A.}},
\bauthor{\bsnm{{Hern{\'a}ndez Su{\'a}rez}}, \binits{E.}},
\bauthor{\bsnm{{Hidalgo}}, \binits{S.L.}},
\bauthor{\bsnm{{Hill}}, \binits{F.}},
\bauthor{\bsnm{{Hizberger}}, \binits{J.}},
\bauthor{\bsnm{{Hofmeister}}, \binits{S.}},
\bauthor{\bsnm{{J{\"a}gers}}, \binits{A.}},
\bauthor{\bsnm{{Janett}}, \binits{G.}},
\bauthor{\bsnm{{Jarolim}}, \binits{R.}},
\bauthor{\bsnm{{Jess}}, \binits{D.}},
\bauthor{\bsnm{{Jim{\'e}nez Mej{\'\i}as}}, \binits{D.}},
\bauthor{\bsnm{{Jolissaint}}, \binits{L.}},
\bauthor{\bsnm{{Kamlah}}, \binits{R.}},
\bauthor{\bsnm{{Kapit{\'a}n}}, \binits{J.}},
\bauthor{\bsnm{{Ka{\v{s}}parov{\'a}}}, \binits{J.}},
\bauthor{\bsnm{{Keller}}, \binits{C.U.}},
\bauthor{\bsnm{{Kentischer}}, \binits{T.}},
\bauthor{\bsnm{{Kiselman}}, \binits{D.}},
\bauthor{\bsnm{{Kleint}}, \binits{L.}},
\bauthor{\bsnm{{Klvana}}, \binits{M.}},
\bauthor{\bsnm{{Kontogiannis}}, \binits{I.}},
\bauthor{\bsnm{{Krishnappa}}, \binits{N.}},
\bauthor{\bsnm{{Ku{\v{c}}era}}, \binits{A.}},
\bauthor{\bsnm{{Labrosse}}, \binits{N.}},
\bauthor{\bsnm{{Lagg}}, \binits{A.}},
\bauthor{\bsnm{{Landi Degl'Innocenti}}, \binits{E.}},
\bauthor{\bsnm{{Langlois}}, \binits{M.}},
\bauthor{\bsnm{{Lafon}}, \binits{M.}},
\bauthor{\bsnm{{Laforgue}}, \binits{D.}},
\bauthor{\bsnm{{Le Men}}, \binits{C.}},
\bauthor{\bsnm{{Lepori}}, \binits{B.}},
\bauthor{\bsnm{{Lepreti}}, \binits{F.}},
\bauthor{\bsnm{{Lindberg}}, \binits{B.}},
\bauthor{\bsnm{{Lilje}}, \binits{P.B.}},
\bauthor{\bsnm{{L{\'o}pez Ariste}}, \binits{A.}},
\bauthor{\bsnm{{L{\'o}pez Fern{\'a}ndez}}, \binits{V.A.}},
\bauthor{\bsnm{{L{\'o}pez Jim{\'e}nez}}, \binits{A.C.}},
\bauthor{\bsnm{{L{\'o}pez L{\'o}pez}}, \binits{R.}},
\bauthor{\bsnm{{Manso Sainz}}, \binits{R.}},
\bauthor{\bsnm{{Marassi}}, \binits{A.}},
\bauthor{\bsnm{{Marco de la Rosa}}, \binits{J.}},
\bauthor{\bsnm{{Marino}}, \binits{J.}},
\bauthor{\bsnm{{Marrero}}, \binits{J.}},
\bauthor{\bsnm{{Mart{\'\i}n}}, \binits{A.}},
\bauthor{\bsnm{{Mart{\'\i}n G{\'a}lvez}}, \binits{A.}},
\bauthor{\bsnm{{Mart{\'\i}n Hernando}}, \binits{Y.}},
\bauthor{\bsnm{{Masciadri}}, \binits{E.}},
\bauthor{\bsnm{{Mart{\'\i}nez Gonz{\'a}lez}}, \binits{M.}},
\bauthor{\bsnm{{Matta-G{\'o}mez}}, \binits{A.}},
\bauthor{\bsnm{{Mato}}, \binits{A.}},
\bauthor{\bsnm{{Mathioudakis}}, \binits{M.}},
\bauthor{\bsnm{{Matthews}}, \binits{S.}},
\bauthor{\bsnm{{Mein}}, \binits{P.}},
\bauthor{\bsnm{{Merlos Garc{\'\i}a}}, \binits{F.}},
\bauthor{\bsnm{{Moity}}, \binits{J.}},
\bauthor{\bsnm{{Montilla}}, \binits{I.}},
\bauthor{\bsnm{{Molinaro}}, \binits{M.}},
\bauthor{\bsnm{{Molodij}}, \binits{G.}},
\bauthor{\bsnm{{Montoya}}, \binits{L.M.}},
\bauthor{\bsnm{{Munari}}, \binits{M.}},
\bauthor{\bsnm{{Murabito}}, \binits{M.}},
\bauthor{\bsnm{{N{\'u}{\~n}ez Cagigal}}, \binits{M.}},
\bauthor{\bsnm{{Oliviero}}, \binits{M.}},
\bauthor{\bsnm{{Orozco Su{\'a}rez}}, \binits{D.}},
\bauthor{\bsnm{{Ortiz}}, \binits{A.}},
\bauthor{\bsnm{{Padilla-Hern{\'a}ndez}}, \binits{C.}},
\bauthor{\bsnm{{Pa{\'e}z Ma{\~n}{\'a}}}, \binits{E.}},
\bauthor{\bsnm{{Paletou}}, \binits{F.}},
\bauthor{\bsnm{{Pancorbo}}, \binits{J.}},
\bauthor{\bsnm{{Pastor Ca{\~n}edo}}, \binits{A.}},
\bauthor{\bsnm{{Pastor Yabar}}, \binits{A.}},
\bauthor{\bsnm{{Peat}}, \binits{A.W.}},
\bauthor{\bsnm{{Pedichini}}, \binits{F.}},
\bauthor{\bsnm{{Peixinho}}, \binits{N.}},
\bauthor{\bsnm{{Pe{\~n}ate}}, \binits{J.}},
\bauthor{\bsnm{{P{\'e}rez de Taoro}}, \binits{A.}},
\bauthor{\bsnm{{Peter}}, \binits{H.}},
\bauthor{\bsnm{{Petrovay}}, \binits{K.}},
\bauthor{\bsnm{{Piazzesi}}, \binits{R.}},
\bauthor{\bsnm{{Pietropaolo}}, \binits{E.}},
\bauthor{\bsnm{{Pleier}}, \binits{O.}},
\bauthor{\bsnm{{Poedts}}, \binits{S.}},
\bauthor{\bsnm{{P{\"o}tzi}}, \binits{W.}},
\bauthor{\bsnm{{Podladchikova}}, \binits{T.}},
\bauthor{\bsnm{{Prieto}}, \binits{G.}},
\bauthor{\bsnm{{Quintero Nehrkorn}}, \binits{J.}},
\bauthor{\bsnm{{Ramelli}}, \binits{R.}},
\bauthor{\bsnm{{Ramos Sapena}}, \binits{Y.}},
\bauthor{\bsnm{{Rasilla}}, \binits{J.L.}},
\bauthor{\bsnm{{Reardon}}, \binits{K.}},
\bauthor{\bsnm{{Rebolo}}, \binits{R.}},
\bauthor{\bsnm{{Regalado Olivares}}, \binits{S.}},
\bauthor{\bsnm{{Reyes Garc{\'\i}a-Talavera}}, \binits{M.}},
\bauthor{\bsnm{{Riethm{\"u}ller}}, \binits{T.L.}},
\bauthor{\bsnm{{Rimmele}}, \binits{T.}},
\bauthor{\bsnm{{Rodr{\'\i}guez Delgado}}, \binits{H.}},
\bauthor{\bsnm{{Rodr{\'\i}guez Gonz{\'a}lez}}, \binits{N.}},
\bauthor{\bsnm{{Rodr{\'\i}guez-Losada}}, \binits{J.A.}},
\bauthor{\bsnm{{Rodr{\'\i}guez Ramos}}, \binits{L.F.}},
\bauthor{\bsnm{{Romano}}, \binits{P.}},
\bauthor{\bsnm{{Roth}}, \binits{M.}},
\bauthor{\bsnm{{Rouppe van der Voort}}, \binits{L.}},
\bauthor{\bsnm{{Rudawy}}, \binits{P.}},
\bauthor{\bsnm{{Ruiz de Galarreta}}, \binits{C.}},
\bauthor{\bsnm{{Ryb{\'a}k}}, \binits{J.}},
\bauthor{\bsnm{{Salvade}}, \binits{A.}},
\bauthor{\bsnm{{S{\'a}nchez-Capuchino}}, \binits{J.}},
\bauthor{\bsnm{{S{\'a}nchez Rodr{\'\i}guez}}, \binits{M.L.}},
\bauthor{\bsnm{{Sangiorgi}}, \binits{M.}},
\bauthor{\bsnm{{Say{\`e}de}}, \binits{F.}},
\bauthor{\bsnm{{Scharmer}}, \binits{G.}},
\bauthor{\bsnm{{Scheiffelen}}, \binits{T.}},
\bauthor{\bsnm{{Schmidt}}, \binits{W.}},
\bauthor{\bsnm{{Schmieder}}, \binits{B.}},
\bauthor{\bsnm{{Scir{\`e}}}, \binits{C.}},
\bauthor{\bsnm{{Scuderi}}, \binits{S.}},
\bauthor{\bsnm{{Siegel}}, \binits{B.}},
\bauthor{\bsnm{{Sigwarth}}, \binits{M.}},
\bauthor{\bsnm{{Sim{\~o}es}}, \binits{P.J.A.}},
\bauthor{\bsnm{{Snik}}, \binits{F.}},
\bauthor{\bsnm{{Sliepen}}, \binits{G.}},
\bauthor{\bsnm{{Sobotka}}, \binits{M.}},
\bauthor{\bsnm{{Socas-Navarro}}, \binits{H.}},
\bauthor{\bsnm{{Sola La Serna}}, \binits{P.}},
\bauthor{\bsnm{{Solanki}}, \binits{S.K.}},
\bauthor{\bsnm{{Soler Trujillo}}, \binits{M.}},
\bauthor{\bsnm{{Soltau}}, \binits{D.}},
\bauthor{\bsnm{{Sordini}}, \binits{A.}},
\bauthor{\bsnm{{Sosa M{\'e}ndez}}, \binits{A.}},
\bauthor{\bsnm{{Stangalini}}, \binits{M.}},
\bauthor{\bsnm{{Steiner}}, \binits{O.}},
\bauthor{\bsnm{{Stenflo}}, \binits{J.O.}},
\bauthor{\bsnm{{{\v{S}}t{\v{e}}p{\'a}n}}, \binits{J.}},
\bauthor{\bsnm{{Strassmeier}}, \binits{K.G.}},
\bauthor{\bsnm{{Sudar}}, \binits{D.}},
\bauthor{\bsnm{{Suematsu}}, \binits{Y.}},
\bauthor{\bsnm{{S{\"u}tterlin}}, \binits{P.}},
\bauthor{\bsnm{{Tallon}}, \binits{M.}},
\bauthor{\bsnm{{Temmer}}, \binits{M.}},
\bauthor{\bsnm{{Tenegi}}, \binits{F.}},
\bauthor{\bsnm{{Tritschler}}, \binits{A.}},
\bauthor{\bsnm{{Trujillo Bueno}}, \binits{J.}},
\bauthor{\bsnm{{Turchi}}, \binits{A.}},
\bauthor{\bsnm{{Utz}}, \binits{D.}},
\bauthor{\bsnm{{van Harten}}, \binits{G.}},
\bauthor{\bsnm{{van Noort}}, \binits{M.}},
\bauthor{\bsnm{{van Werkhoven}}, \binits{T.}},
\bauthor{\bsnm{{Vansintjan}}, \binits{R.}},
\bauthor{\bsnm{{Vaz Cedillo}}, \binits{J.J.}},
\bauthor{\bsnm{{Vega Reyes}}, \binits{N.}},
\bauthor{\bsnm{{Verma}}, \binits{M.}},
\bauthor{\bsnm{{Veronig}}, \binits{A.M.}},
\bauthor{\bsnm{{Viavattene}}, \binits{G.}},
\bauthor{\bsnm{{Vitas}}, \binits{N.}},
\bauthor{\bsnm{{V{\"o}gler}}, \binits{A.}},
\bauthor{\bsnm{{von der L{\"u}he}}, \binits{O.}},
\bauthor{\bsnm{{Volkmer}}, \binits{R.}},
\bauthor{\bsnm{{Waldmann}}, \binits{T.A.}},
\bauthor{\bsnm{{Walton}}, \binits{D.}},
\bauthor{\bsnm{{Wisniewska}}, \binits{A.}},
\bauthor{\bsnm{{Zeman}}, \binits{J.}},
\bauthor{\bsnm{{Zeuner}}, \binits{F.}},
\bauthor{\bsnm{{Zhang}}, \binits{L.Q.}},
\bauthor{\bsnm{{Zuccarello}}, \binits{F.}},
\bauthor{\bsnm{{Collados}}, \binits{M.}}:
\byear{2022},
\batitle{{The European Solar Telescope}}.
\bjtitle{\aap}
\bvolume{666},
\bfpage{A21}.
\doiurl{https://doi.org/10.1051/0004-6361/202243867}.
\adsurl{2022A&A...666A..21Q}.
\end{barticle}
\endbibitem

\bibitem[\protect\citeauthoryear{{Rimmele} et~al.}{2020}]{Rimmele2020}
\begin{barticle}
\bauthor{\bsnm{{Rimmele}}, \binits{T.R.}},
\bauthor{\bsnm{{Warner}}, \binits{M.}},
\bauthor{\bsnm{{Keil}}, \binits{S.L.}},
\bauthor{\bsnm{{Goode}}, \binits{P.R.}},
\bauthor{\bsnm{{Kn{\"o}lker}}, \binits{M.}},
\bauthor{\bsnm{{Kuhn}}, \binits{J.R.}},
\bauthor{\bsnm{{Rosner}}, \binits{R.R.}},
\bauthor{\bsnm{{McMullin}}, \binits{J.P.}},
\bauthor{\bsnm{{Casini}}, \binits{R.}},
\bauthor{\bsnm{{Lin}}, \binits{H.}},
\bauthor{\bsnm{{W{\"o}ger}}, \binits{F.}},
\bauthor{\bsnm{{von der L{\"u}he}}, \binits{O.}},
\bauthor{\bsnm{{Tritschler}}, \binits{A.}},
\bauthor{\bsnm{{Davey}}, \binits{A.}},
\bauthor{\bsnm{{de Wijn}}, \binits{A.}},
\bauthor{\bsnm{{Elmore}}, \binits{D.F.}},
\bauthor{\bsnm{{Fehlmann}}, \binits{A.}},
\bauthor{\bsnm{{Harrington}}, \binits{D.M.}},
\bauthor{\bsnm{{Jaeggli}}, \binits{S.A.}},
\bauthor{\bsnm{{Rast}}, \binits{M.P.}},
\bauthor{\bsnm{{Schad}}, \binits{T.A.}},
\bauthor{\bsnm{{Schmidt}}, \binits{W.}},
\bauthor{\bsnm{{Mathioudakis}}, \binits{M.}},
\bauthor{\bsnm{{Mickey}}, \binits{D.L.}},
\bauthor{\bsnm{{Anan}}, \binits{T.}},
\bauthor{\bsnm{{Beck}}, \binits{C.}},
\bauthor{\bsnm{{Marshall}}, \binits{H.K.}},
\bauthor{\bsnm{{Jeffers}}, \binits{P.F.}},
\bauthor{\bsnm{{Oschmann}}, \binits{J.M.}},
\bauthor{\bsnm{{Beard}}, \binits{A.}},
\bauthor{\bsnm{{Berst}}, \binits{D.C.}},
\bauthor{\bsnm{{Cowan}}, \binits{B.A.}},
\bauthor{\bsnm{{Craig}}, \binits{S.C.}},
\bauthor{\bsnm{{Cross}}, \binits{E.}},
\bauthor{\bsnm{{Cummings}}, \binits{B.K.}},
\bauthor{\bsnm{{Donnelly}}, \binits{C.}},
\bauthor{\bsnm{{de Vanssay}}, \binits{J.-B.}},
\bauthor{\bsnm{{Eigenbrot}}, \binits{A.D.}},
\bauthor{\bsnm{{Ferayorni}}, \binits{A.}},
\bauthor{\bsnm{{Foster}}, \binits{C.}},
\bauthor{\bsnm{{Galapon}}, \binits{C.A.}},
\bauthor{\bsnm{{Gedrites}}, \binits{C.}},
\bauthor{\bsnm{{Gonzales}}, \binits{K.}},
\bauthor{\bsnm{{Goodrich}}, \binits{B.D.}},
\bauthor{\bsnm{{Gregory}}, \binits{B.S.}},
\bauthor{\bsnm{{Guzman}}, \binits{S.S.}},
\bauthor{\bsnm{{Guzzo}}, \binits{S.}},
\bauthor{\bsnm{{Hegwer}}, \binits{S.}},
\bauthor{\bsnm{{Hubbard}}, \binits{R.P.}},
\bauthor{\bsnm{{Hubbard}}, \binits{J.R.}},
\bauthor{\bsnm{{Johansson}}, \binits{E.M.}},
\bauthor{\bsnm{{Johnson}}, \binits{L.C.}},
\bauthor{\bsnm{{Liang}}, \binits{C.}},
\bauthor{\bsnm{{Liang}}, \binits{M.}},
\bauthor{\bsnm{{McQuillen}}, \binits{I.}},
\bauthor{\bsnm{{Mayer}}, \binits{C.}},
\bauthor{\bsnm{{Newman}}, \binits{K.}},
\bauthor{\bsnm{{Onodera}}, \binits{B.}},
\bauthor{\bsnm{{Phelps}}, \binits{L.}},
\bauthor{\bsnm{{Puentes}}, \binits{M.M.}},
\bauthor{\bsnm{{Richards}}, \binits{C.}},
\bauthor{\bsnm{{Rimmele}}, \binits{L.M.}},
\bauthor{\bsnm{{Sekulic}}, \binits{P.}},
\bauthor{\bsnm{{Shimko}}, \binits{S.R.}},
\bauthor{\bsnm{{Simison}}, \binits{B.E.}},
\bauthor{\bsnm{{Smith}}, \binits{B.}},
\bauthor{\bsnm{{Starman}}, \binits{E.}},
\bauthor{\bsnm{{Sueoka}}, \binits{S.R.}},
\bauthor{\bsnm{{Summers}}, \binits{R.T.}},
\bauthor{\bsnm{{Szabo}}, \binits{A.}},
\bauthor{\bsnm{{Szabo}}, \binits{L.}},
\bauthor{\bsnm{{Wampler}}, \binits{S.B.}},
\bauthor{\bsnm{{Williams}}, \binits{T.R.}},
\bauthor{\bsnm{{White}}, \binits{C.}}:
\byear{2020},
\batitle{{The Daniel K. Inouye Solar Telescope - Observatory Overview}}.
\bjtitle{\solphys}
\bvolume{295},
\bfpage{172}.
\doiurl{https://doi.org/10.1007/s11207-020-01736-7}.
\adsurl{2020SoPh..295..172R}.
\end{barticle}
\endbibitem

\bibitem[\protect\citeauthoryear{{Scharmer} et~al.}{2008}]{Scharmer2008}
\begin{barticle}
\bauthor{\bsnm{{Scharmer}}, \binits{G.B.}},
\bauthor{\bsnm{{Narayan}}, \binits{G.}},
\bauthor{\bsnm{{Hillberg}}, \binits{T.}},
\bauthor{\bsnm{{de la Cruz Rodriguez}}, \binits{J.}},
\bauthor{\bsnm{{L{\"o}fdahl}}, \binits{M.G.}},
\bauthor{\bsnm{{Kiselman}}, \binits{D.}},
\bauthor{\bsnm{{S{\"u}tterlin}}, \binits{P.}},
\bauthor{\bsnm{{van Noort}}, \binits{M.}},
\bauthor{\bsnm{{Lagg}}, \binits{A.}}:
\byear{2008},
\batitle{{CRISP Spectropolarimetric Imaging of Penumbral Fine Structure}}.
\bjtitle{\apjl}
\bvolume{689},
\bfpage{L69}.
\doiurl{https://doi.org/10.1086/595744}.
\adsurl{2008ApJ...689L..69S}.
\end{barticle}
\endbibitem

\bibitem[\protect\citeauthoryear{{Schou} et~al.}{2012}]{Schou2012}
\begin{barticle}
\bauthor{\bsnm{{Schou}}, \binits{J.}},
\bauthor{\bsnm{{Scherrer}}, \binits{P.H.}},
\bauthor{\bsnm{{Bush}}, \binits{R.I.}},
\bauthor{\bsnm{{Wachter}}, \binits{R.}},
\bauthor{\bsnm{{Couvidat}}, \binits{S.}},
\bauthor{\bsnm{{Rabello-Soares}}, \binits{M.C.}},
\bauthor{\bsnm{{Bogart}}, \binits{R.S.}},
\bauthor{\bsnm{{Hoeksema}}, \binits{J.T.}},
\bauthor{\bsnm{{Liu}}, \binits{Y.}},
\bauthor{\bsnm{{Duvall}}, \binits{T.L.}},
\bauthor{\bsnm{{Akin}}, \binits{D.J.}},
\bauthor{\bsnm{{Allard}}, \binits{B.A.}},
\bauthor{\bsnm{{Miles}}, \binits{J.W.}},
\bauthor{\bsnm{{Rairden}}, \binits{R.}},
\bauthor{\bsnm{{Shine}}, \binits{R.A.}},
\bauthor{\bsnm{{Tarbell}}, \binits{T.D.}},
\bauthor{\bsnm{{Title}}, \binits{A.M.}},
\bauthor{\bsnm{{Wolfson}}, \binits{C.J.}},
\bauthor{\bsnm{{Elmore}}, \binits{D.F.}},
\bauthor{\bsnm{{Norton}}, \binits{A.A.}},
\bauthor{\bsnm{{Tomczyk}}, \binits{S.}}:
\byear{2012},
\batitle{{Design and Ground Calibration of the Helioseismic and Magnetic Imager
  (HMI) Instrument on the Solar Dynamics Observatory (SDO)}}.
\bjtitle{\solphys}
\bvolume{275},
\bfpage{229}.
\doiurl{https://doi.org/10.1007/s11207-011-9842-2}.
\adsurl{2012SoPh..275..229S}.
\end{barticle}
\endbibitem

\bibitem[\protect\citeauthoryear{{Semel}}{1967}]{Semel1967}
\begin{barticle}
\bauthor{\bsnm{{Semel}}, \binits{M.}}:
\byear{1967},
\batitle{{Contribution {\`a} l{\'e}tude des champs magn{\'e}tiques dans les
  r{\'e}gions actives solaires}}.
\bjtitle{Annales d'Astrophysique}
\bvolume{30},
\bfpage{513}.
\adsurl{1967AnAp...30..513S}.
\end{barticle}
\endbibitem

\bibitem[\protect\citeauthoryear{{Severny}}{1955}]{Severny1955}
\begin{barticle}
\bauthor{\bsnm{{Severny}}, \binits{A.B.}}:
\byear{1955},
\batitle{{The Solar Tower telescope at the Crimean Astrophysical Observatory}}.
\bjtitle{Izv. Krymsk. Astrofiz. Observ. (in Russian)}
\bvolume{15},
\bfpage{31}.
\end{barticle}
\endbibitem

\bibitem[\protect\citeauthoryear{{Skumanich} et~al.}{1997}]{Skumanich1997}
\begin{barticle}
\bauthor{\bsnm{{Skumanich}}, \binits{A.}},
\bauthor{\bsnm{{Lites}}, \binits{B.W.}},
\bauthor{\bsnm{{Mart{\'\i}nez Pillet}}, \binits{V.}},
\bauthor{\bsnm{{Seagraves}}, \binits{P.}}:
\byear{1997},
\batitle{{The Calibration of the Advanced Stokes Polarimeter}}.
\bjtitle{\apjs}
\bvolume{110},
\bfpage{357}.
\doiurl{https://doi.org/10.1086/313004}.
\adsurl{1997ApJS..110..357S}.
\end{barticle}
\endbibitem

\bibitem[\protect\citeauthoryear{{Smitha} and {Solanki}}{2017}]{Smitha2017}
\begin{barticle}
\bauthor{\bsnm{{Smitha}}, \binits{H.N.}},
\bauthor{\bsnm{{Solanki}}, \binits{S.K.}}:
\byear{2017},
\batitle{{Probing photospheric magnetic fields with new spectral line pairs}}.
\bjtitle{\aap}
\bvolume{608},
\bfpage{A111}.
\doiurl{https://doi.org/10.1051/0004-6361/201731261}.
\adsurl{2017A&A...608A.111S}.
\end{barticle}
\endbibitem

\bibitem[\protect\citeauthoryear{{Varsik} and {Yang}}{2006}]{Varsik2006}
\begin{bchapter}
\bauthor{\bsnm{{Varsik}}, \binits{J.R.}},
\bauthor{\bsnm{{Yang}}, \binits{G.}}:
\byear{2006},
\bctitle{{Design of a telescope pointing and tracking subsystem for the Big
  Bear Solar Observatory New Solar Telescope}}.
In: \beditor{\bsnm{{Lewis}}, \binits{H.}},
\beditor{\bsnm{{Bridger}}, \binits{A.}} (eds.)
\bbtitle{Advanced Software and Control for Astronomy},
\bsertitle{Society of Photo-Optical Instrumentation Engineers (SPIE) Conference
  Series}
\bseriesno{6274},
\bfpage{62741T}.
\doiurl{https://doi.org/10.1117/12.672039}.
\adsurl{2006SPIE.6274E..1TV}.
\end{bchapter}
\endbibitem

\bibitem[\protect\citeauthoryear{{Xu} et~al.}{2024}]{Xu2024}
\begin{barticle}
\bauthor{\bsnm{{Xu}}, \binits{H.}},
\bauthor{\bsnm{{Su}}, \binits{J.}},
\bauthor{\bsnm{{Liu}}, \binits{S.}},
\bauthor{\bsnm{{Deng}}, \binits{Y.}},
\bauthor{\bsnm{{Yang}}, \binits{S.}},
\bauthor{\bsnm{{Bai}}, \binits{X.}},
\bauthor{\bsnm{{Chen}}, \binits{J.}},
\bauthor{\bsnm{{Wang}}, \binits{X.}},
\bauthor{\bsnm{{Yang}}, \binits{X.}},
\bauthor{\bsnm{{Song}}, \binits{Y.}},
\bauthor{\bsnm{{Idrees}}, \binits{S.}}:
\byear{2024},
\batitle{{Comparison of Line-of-Sight Magnetic Field Observed by ASO-S/FMG,
  SDO/HMI and HSOS/SMAT}}.
\bjtitle{\solphys}
\bvolume{299},
\bfpage{17}.
\doiurl{https://doi.org/10.1007/s11207-024-02260-8}.
\adsurl{2024SoPh..299...17X}.
\end{barticle}
\endbibitem

\bibitem[\protect\citeauthoryear{{Yang} et~al.}{2006}]{Yang2006}
\begin{bchapter}
\bauthor{\bsnm{{Yang}}, \binits{G.}},
\bauthor{\bsnm{{Varsik}}, \binits{J.R.}},
\bauthor{\bsnm{{Shumko}}, \binits{S.}},
\bauthor{\bsnm{{Denker}}, \binits{C.}},
\bauthor{\bsnm{{Choi}}, \binits{S.}},
\bauthor{\bsnm{{Verdoni}}, \binits{A.P.}},
\bauthor{\bsnm{{Wang}}, \binits{H.}}:
\byear{2006},
\bctitle{{The telescope control system of the New Solar Telescope at Big Bear
  Solar Observatory}}.
In: \beditor{\bsnm{{Lewis}}, \binits{H.}},
\beditor{\bsnm{{Bridger}}, \binits{A.}} (eds.)
\bbtitle{Advanced Software and Control for Astronomy},
\bsertitle{Society of Photo-Optical Instrumentation Engineers (SPIE) Conference
  Series}
\bseriesno{6274},
\bfpage{62741Y}.
\doiurl{https://doi.org/10.1117/12.672402}.
\adsurl{2006SPIE.6274E..1YY}.
\end{bchapter}
\endbibitem

\end{thebibliography}

\IfFileExists{\jobname.bbl}{} {\typeout{}
\typeout{****************************************************}
\typeout{****************************************************}
\typeout{** Please run "bibtex \jobname" to obtain} \typeout{**
the bibliography and then re-run LaTeX} \typeout{** twice to fix
the references !}
\typeout{****************************************************}
\typeout{****************************************************}
\typeout{}}

\end{document}